\documentclass[a4paper, 12pt, oneside]{article}
\pdfoutput=1
\usepackage{jheppub}
\makeatletter
\def\@fpheader{}
\makeatother

\usepackage{graphicx,wrapfig,float,slashed,subcaption}
\usepackage{amsmath,amssymb,epsfig,graphicx,xcolor}
\usepackage{epstopdf}
\usepackage{multirow}

\usepackage{lscape} 

\usepackage{ragged2e}
\usepackage{lipsum}
\usepackage{array}

\usepackage{soul}

\usepackage{color}

\usepackage{tikz}
\usetikzlibrary{arrows,positioning,shapes}
\usetikzlibrary{decorations.pathmorphing}
\usetikzlibrary{decorations.markings}

\usepackage{cleveref}

\newcommand{\htb}[1]{{\color{blue} #1}}

\newcommand{\nc}{\newcommand}
\nc{\non}{\nonumber}
\nc{\hc}{\hbox {H.c.}}
\nc{\noi}{\noinde	nt}
\nc{\barx}{\bar{x}}
\nc{\pbarn}{\;\hbox {pb}}
\nc{\fbarn}{\;\hbox {fb}}
\nc{\ra}{\rightarrow}
\nc{\met}{p_{T}^{\textnormal{miss}}}
\nc{\lep}{\ell}
\nc{\pb}{\;\hbox {pb}}

\nc{\lb}{\left(}
\nc{\rb}{\right)}
\usepackage{color}

\definecolor{agray}{rgb}{0.95, 0.95, 0.99}
\nc{\hsp}{\hspace{0.5cm}}
\nc{\lsp}{\hspace{1cm}}
\nc{\Lsp}{\hspace{2cm}}
\nc{\LLsp}{\lsp\lsp}
\nc{\lra}{\longrightarrow}
\nc{\p}{\prime}
\nc{\sgn}{\text{sgn}}
\nc{\ph}{\varphi}
\newcommand{\lam}{\lambda}
\nc{\beq}{\begin{equation}}  \nc{\eeq}{\end{equation}}
\nc{\bea}{\begin{eqnarray}}  \nc{\eea}{\end{eqnarray}}
\nc{\baa}{\begin{array}}     \nc{\eaa}{\end{array}}
\nc{\bit}{\begin{itemize}}   \nc{\eit}{\end{itemize}}
\nc{\ben}{\begin{enumerate}} \nc{\een}{\end{enumerate}}
\nc{\bce}{\begin{center}}    \nc{\ece}{\end{center}}
\nc{\bpm}{\begin{pmatrix}}   \nc{\epm}{\end{pmatrix}}
\nc{\bvt}{\begin{verbatim}}  \nc{\evt}{\end{verbatim}}

\def\lsim{\mathrel{\raise.3ex\hbox{$<$\kern-.75em\lower1ex\hbox{$\sim$}}}}
\def\gsim{\mathrel{\raise.3ex\hbox{$>$\kern-.75em\lower1ex\hbox{$\sim$}}}}

\def\udots{\mathinner{\mkern1mu\raise1pt\vbox{\kern7pt\hbox{.}}\mkern2mu\raise4pt\hbox{.}\mkern2mu\raise7pt\hbox{.}\mkern1mu}}

\usepackage{xspace}

\newcommand{\TeV}{\ensuremath{\mathrm{\,Te\kern -0.1em V}}\xspace}
\newcommand{\GeV}{\ensuremath{\mathrm{\,Ge\kern -0.1em V}}\xspace}
\newcommand{\tev}{\ensuremath{\mathrm{\,Te\kern -0.1em V}}\xspace}
\newcommand{\gev}{\ensuremath{\mathrm{\,Ge\kern -0.1em V}}\xspace}
\newcommand{\mev}{\ensuremath{\mathrm{\,Me\kern -0.1em V}}\xspace}

\newcommand{\fb}{\ensuremath{\,\text{fb}}\xspace}
\newcommand{\ab}{\ensuremath{\,\text{ab}}\xspace}
\newcommand{\abinv}{\ensuremath{\,\text{ab}^{-1}}\xspace}

\newcommand{\HS}{\texttt{HiggsSignals}}
\newcommand{\HB}{\texttt{HiggsBounds}}

\newcommand\fverb{\setbox\fverbbox=\hbox\bgroup\verb}
\newcommand\fverbdo{\egroup\medskip\noindent%
			\fbox{\unhbox\fverbbox}\ }
\newcommand\fverbit{\egroup\item[\fbox{\unhbox\fverbbox}]}
\newbox\fverbbox

\makeatletter
\renewcommand{\boxed}[1]{\textcolor{black}{%
\tikz[baseline={([yshift=-0ex]current bounding box.center)}] \node [rectangle, minimum width=0ex,draw] {\normalcolor\m@th$\displaystyle#1$};}}
 \makeatother

\title{Exploring Inert Scalars at CLIC}

\author[a]{Jan Kalinowski,}
\author[b]{Wojciech Kotlarski,}
\author[c,d]{Tania Robens,}
\author[a,e]{Dorota Soko\l owska,}
\author[a]{Aleksander~Filip~\.Zarnecki}
\affiliation[a]{Faculty of Physics, University of Warsaw, Warsaw, Poland}
\affiliation[b] {Institut f\"ur Kern- und Teilchenphysik, TU Dresden, Dresden, Germany}
\affiliation[c]{MTA-DE Particle Physics Research Group, University of
  Debrecen, Debrecen, Hungary}
\affiliation[d]{Theoretical Physics Division, Rudjer Boskovic Institute, 10002 Zagreb, Croatia}
\affiliation[e]{International Institute of Physics, Universidade Federal do Rio Grande do Norte, 
Campus Universitario, Lagoa Nova, Natal-RN 59078-970, Brazil}
\emailAdd{jan.kalinowski@fuw.edu.pl}
\emailAdd{wojciech.kotlarski@tu-dresden.de}
\emailAdd{trobens@irb.hr}
\emailAdd{dsokolowska@iip.ufrn.br}
\emailAdd{filip.zarnecki@fuw.edu.pl}

\abstract{%
  We investigate the prospect of discovering
  the Inert Doublet Model scalars at CLIC.
  As signal processes,   we consider the pair-production of
  inert scalars, namely  
  $e^{+}e^{-} \rightarrow  H^{+}H^{-}$ and $e^{+}e^{-}\rightarrow AH$,
  followed by decays of charged scalars $H^\pm$ and  neutral scalars $A$
  into leptonic final states and missing transverse energy.

  We focus on signal signatures with two muons or an
  electron and a muon pair in the final state.  A number of selected
  benchmark scenarios that cover the range of possible collider
  signatures of the IDM are considered. For the suppression of SM background with the
  same visible signature, multivariate
  analysis methods are employed. For several benchmark points discovery is
  already possible at {the} low-energy stage of CLIC. 
  Prospects of investigating scenarios that are only accessible at
  higher collider energies are also discussed.
 }


\tikzstyle{every picture}+=[remember picture]
\pgfdeclaredecoration{complete sines}{initial}
{
    \state{initial}[
        width=+0pt,
        next state=sine,
        persistent precomputation={\pgfmathsetmacro\matchinglength{
            \pgfdecoratedinputsegmentlength / int(\pgfdecoratedinputsegmentlength/\pgfdecorationsegmentlength)}
            \setlength{\pgfdecorationsegmentlength}{\matchinglength pt}
        }] {}
    \state{sine}[width=\pgfdecorationsegmentlength]{
        \pgfpathsine{\pgfpoint{0.25\pgfdecorationsegmentlength}{0.5\pgfdecorationsegmentamplitude}}
        \pgfpathcosine{\pgfpoint{0.25\pgfdecorationsegmentlength}{-0.5\pgfdecorationsegmentamplitude}}
        \pgfpathsine{\pgfpoint{0.25\pgfdecorationsegmentlength}{-0.5\pgfdecorationsegmentamplitude}}
        \pgfpathcosine{\pgfpoint{0.25\pgfdecorationsegmentlength}{0.5\pgfdecorationsegmentamplitude}}
}
    \state{final}{}
}

\tikzset{
fermion/.style={very thick,draw=black, line cap=round, postaction={decorate},
    decoration={markings,mark=at position .65 with {\arrow[black]{latex}}}},
photon/.style={very thick, line cap=round,decorate, draw=black,
    decoration={complete sines,amplitude=4pt, segment length=8pt}},
boson/.style={very thick, line cap=round,decorate, draw=black,
    decoration={complete sines,amplitude=4pt,segment length=8pt}},
gluon/.style={very thick,line cap=round, decorate, draw=black,
    decoration={coil,aspect=1,amplitude=4pt, segment length=8pt}},
scalar/.style={dashed, very thick,line cap=round, decorate, draw=black},
cscalar/.style={dashed,very  thick,line cap=round, draw=black,postaction={decorate},
    decoration={markings,mark=at position .65 with {\arrow[black]{latex}}}},
jet/.style={double, thick,line cap=round, decorate, draw=black},
ghost/.style={dotted, thick,line cap=round, decorate, draw=black},
->-/.style={decoration={
  markings,
  mark=at position 0.6 with {\arrow{>}}},postaction={decorate}}
 }

 \makeatletter
\tikzset{
    position/.style args={#1 degrees from #2}{
        at=(#2.#1), anchor=#1+180, shift=(#1:\tikz@node@distance)
    }
}
\makeatother

\notoc

\renewcommand\afterLogoSpace{\vspace*{0.5cm}}     

\renewcommand\afterEmailSpace{\vskip36pt plus9pt minus10pt\filbreak}

\begin{document}
\begin{flushright}
DMIIP-2018
\end{flushright}
\maketitle


\newpage



\section{Introduction}
\label{Introduction}

A number of astrophysical observations based on gravitational
interactions point to the existence of dark matter (DM) in the
Universe, which can not be described with the Standard Model (SM).
One of the simplest extensions of the SM, which can provide a dark
matter candidate is the Inert Doublet Model (IDM)
\cite{Deshpande:1977rw,Cao:2007rm,Barbieri:2006dq}.
The scalar sector of the IDM consists of two SU(2) doublets where one is
the SM-like Higgs doublet, $\Phi_S$, while the other is called inert
or dark doublet, $\Phi_D$.  After electroweak symmetry breaking the sector has five
physical states: apart from the SM Higgs boson $h$ it has two  neutral
ones, $H$ and $A$, as well as two charged scalars, $H^\pm$. {A discrete $Z_2$ symmetry prohibits the inert scalars from interacting
with the SM fermions through Yukawa-type interactions and makes the
lightest neutral scalar, chosen to be $H$ in this work, a good dark
matter candidate.}  

In this work we study the potential of CLIC running at three energy
stages {as a discovery machine for the IDM scalars}.
We consider neutral scalar ($H A$) and charged scalar
($H^+H^-$) pair-production at center-of-mass energies of 380\GeV,
1.5\TeV and 3\TeV.  
We explore a set of benchmark points (BPs) proposed in \cite{Kalinowski:2018ylg},
which satisfy all the recent experimental and theoretical constraints,
provide the neutral $H$ boson as the dark matter candidate
($m_H < m_{H^\pm}, m_A$), and span the inert scalar mass range from
about 50\GeV to 1\TeV. 
Earlier analyses of the IDM at colliders were done in
\cite{LopezHonorez:2006gr,Cao:2007rm,Lundstrom:2008ai,Dolle:2009ft,Dolle:2009fn,Honorez:2010re,Miao:2010rg,Gustafsson:2012aj,Arhrib:2012ia,Swiezewska:2012eh,Aoki:2013lhm,Ho:2013spa,Arhrib:2013ela,Krawczyk:2013jta,Goudelis:2013uca,Ginzburg:2014ora,Belanger:2015kga,Blinov:2015qva,Arhrib:2015hoa,Ilnicka:2015jba,Poulose:2016lvz,Datta:2016nfz,Kanemura:2016sos,Akeroyd:2016ymd,Wan:2018eaz,Ilnicka:2018def,Belyaev:2018ext}.

The paper is organized as follows. The structure of the IDM scalar sector 
and the considered benchmark points are briefly described in \cref{Model}.
In \cref{Simulation} the analysis strategy and simulation tools are
described. In \cref{Low energy} results on the possible measurement
of low-mass benchmark points at the first stage of CLIC, at 380 \GeV,
are presented, while \cref{High energy} discusses the 
prospects for heavy inert scalar production at high-energy CLIC.
Finally, the conclusions are given in \cref{Conclusion}.



\section{Inert Doublet Model}
\label{Model}

The scalar sector of the IDM consists of two scalar doublets, the SM Higgs
doublet $\Phi_S$  with SM-like Higgs boson $h$ and the inert doublet
$\Phi_D$ with four inert scalars $H,\,A,\,H^\pm$.
A discrete $Z_2$ symmetry is imposed under which the SM-like Higgs
doublet $\Phi_S$ and all the other SM fields are {\it even}, whereas
the inert doublet $\Phi_D$ is {\it odd}.
As a result, inert scalars do not interact with the SM fermions through
Yukawa-type interactions, and the 
most general renormalizable scalar potential for the IDM 
{is given by

\begin{equation}\begin{array}{c}
V=-\frac{1}{2}\left[m_{11}^2(\Phi_S^\dagger\Phi_S)\!+\! m_{22}^2(\Phi_D^\dagger\Phi_D)\right]+
\frac{\lambda_1}{2}(\Phi_S^\dagger\Phi_S)^2\! 
+\!\frac{\lambda_2}{2}(\Phi_D^\dagger\Phi_D)^2\\[2mm]+\!\lambda_3(\Phi_S^\dagger\Phi_S)(\Phi_D^\dagger\Phi_D)\!
\!+\!\lambda_4(\Phi_S^\dagger\Phi_D)(\Phi_D^\dagger\Phi_S) +\frac{\lambda_5}{2}\left[(\Phi_S^\dagger\Phi_D)^2\!
+\!(\Phi_D^\dagger\Phi_S)^2\right].
\end{array}\label{pot}\end{equation}

}

{A more detailed discussion of the potential and physical parameters can be found in section 2 of \cite{Ilnicka:2015jba}.}

Due to the exact $Z_2$ symmetry, the lightest neutral scalar $H$ (or
$A$) is stable and thereby may serve as a good dark matter
candidate. In this work we choose $H$ to be the lightest particle (choosing $A$ instead of $H$ as the 
lightest particle changes only the meaning of $\lambda_5\to -\lambda_5$).
After fixing the SM-like Higgs boson mass to $m_h=125.1$\,\GeV~  and the
vacuum expectation value of the SM-like doublet to $v=246$\,\GeV~ (the
SM value) we are left with 5 free parameters, which we take as
\begin{equation}\label{phys_set}
m_H , m_A , m_{H^\pm} , \lambda_2 , \lambda_{345},
\end{equation}
where $ \lambda_{345}=\lambda_3+\lambda_4+\lambda_5$ determines the
Higgs-DM coupling, while $\lambda_2$ corresponds to couplings within
the dark sector.  

To study the prospects of IDM scalar measurement at CLIC we consider a
set of benchmark points proposed in \cite{Kalinowski:2018ylg}, and listed in
\cref{tab:bench} for the low-mass benchmarks accessible at 380 \GeV, and in
\cref{tab:bench2} for the high-mass
benchmarks accessible only at higher collider energies of 1.5 and 3 \TeV.
These benchmarks were selected from a larger set of points in the IDM
parameter space, which were found to be in agreement with all the
theoretical and current experimental constraints.
Points corresponding to different assignment of masses and couplings
were selected in the parameter range interesting in view of future linear $e^+e^-$ collider searches.

We refer the reader to \cite{Ilnicka:2015jba,Ilnicka:2018def,Kalinowski:2018ylg} for a detailed discussion of the theoretical and experimental
constraints and the benchmark selection; comments on the impact of future XENON-nT measurements and prospects of testing the IDM model at the LHC can be found in~\cite{Kalinowski:2018ylg}. 
{We summarize the crucial
  points of the discussion presented there below. 
\paragraph{Theoretical and experimental constraints} 
As theoretical constraints,   positivity of the
potential \cite{Nie:1998yn}, the condition to be in the global
  inert vacuum \cite{Ginzburg:2010wa}  and
perturbative unitarity \cite{Chanowitz:1985hj,Ginzburg:2005dt}
have been imposed. We
  also require
agreement with electroweak precision tests \cite{Baak:2014ora} via
oblique parameters
\cite{Altarelli:1990zd,Peskin:1990zt,Peskin:1991sw,Maksymyk:1993zm},
{zero} contributions to electroweak gauge boson widths from inert
particles {by kinematically forbidding decays $W^\pm \to A H^\pm, H
  H^\pm, Z \to AH, H^+H^-$} \cite{Tanabashi:2018oca},  and agreement with recasts of LEP and LHC
searches \cite{Lundstrom:2008ai,Belanger:2015kga}, including a
  lower limit of 70 \GeV for the charged scalar mass \cite{Pierce:2007ut}\footnote{{In recent 
  work \cite{Dercks:2018wch}, an additional
    recast has been presented which uses 13 \TeV LHC data. We found
    that our benchmark points are not constrained by the limits
    derived in that reference.}}. We also set a hard upper cutoff on the charged scalar
  life-time to avoid constraints from long-lived charged particle
  searches; a more detailed study of these constraints has recently
  been presented in \cite{Heisig:2018kfq}. We make use
  of recent LHC findings to constrain the total width of the Higgs
  particle \cite{Sirunyan:2019twz}, its invisible branching ratio
  \cite{Khachatryan:2016whc} and the branching ratio
  $h\,\rightarrow\,\gamma\gamma$
  \cite{Khachatryan:2016vau}\footnote{{Additional tests of agreement
    with collider findings were performed using \HB~\cite{Bechtle:2008jh,Bechtle:2011sb,Bechtle:2013gu,Bechtle:2013wla,Bechtle:2015pma} and \HS~\cite{Bechtle:2013xfa,Bechtle:2014ewa}.}}. In order to not
  overclose the universe the relic density of
dark matter candidate $H$ was required to be at most within a
two sigma range of the value recently published by the Planck
experiment, i.e. $\Omega_c\,h^2\,\leq\,0.1224$
\cite{Aghanim:2018eyx}. Relic density has been calculated including all relevant
channels, in particular decays into virtual gauge bosons. The
  proposed benchmarks have been selected (out of 13500 that passed
  all above criteria) to represent different signatures at $e^+e^-$
  colliders that differ $e.g.$ in the mass spectrum or
    production cross sections, in order to cover a wide variety of
    possible collider signals. For the selected benchmark
points, all of which correspond to masses of DM particles below 1 TeV,
the Sommerfeld enhancement does not play a relevant role
\cite{Garcia-Cely:2015khw}. Finally the 
agreement with results from direct detection experiments
\cite{Aprile:2018dbl} has been
  required \footnote{{Electroweak higher-order corrections to
      direct detection cross section within the IDM have been
      presented in \cite{Klasen:2013btp}. From that work, we can
      estimate the one-loop contributions for out benchmark points to
      be $\mathcal{O}\lb 10^{-11} \pb \rb$. We explicitly checked
      that even assuming such a maximal additional contribution does
      not exclude any of our BPs.}}.
As discussed in \cite{Kalinowski:2018ylg}, constraints from
 indirect detection are weaker than direct detection constraints
    (see also e.g. discussion in \cite{Aartsen:2016zhm}). The most
 stringent limits stem from indirect detection leading to $b\,\bar{b}$
 final states \cite{Ahnen:2016qkx}.   
These can be easily
  avoided by tuning the value of $\lam_{345}$, which is
  irrelevant for the collider phenomenology discussed here. It should
  also be noted that dark matter predictions for some points, especially
  BP23, highly depend on the input in the SM mass spectrum;
  e.g. variations within 3 standard deviations for bottom and Higgs
  masses can lead to large variations in the predictions (up to a
  factor 5), while these are of no importance for our studies.

In the scan presented in \cite{Kalinowski:2018ylg},
we made use of the publicly available tools: the \texttt{2HDMC} code
\cite{Eriksson:2009ws} and \texttt{micrOMEGAs} version 4.3.5
\cite{Barducci:2016pcb} to apply several of the constraints discussed above.

\paragraph{Prospects for detection at the
    LHC}
In general, the IDM can 
be tested at the LHC through a variety of signatures, including
mono-jet, mono-Z, mono-Higgs and Vector-Boson-Fusion + missing
transverse energy signatures, as well as through multi-lepton and
multi-jet final states, see discussions in
\cite{Poulose:2016lvz,Datta:2016nfz,Wan:2018eaz,Datta:2016nfz,Dutta:2017lny,Wan:2018eaz}. In
\cite{Dercks:2018wch}, a dedicated discussion suggests that current
searches, e.g. for multi-lepton final states, would have to be
significantly modified in order to access the current IDM parameter
space even for low dark matter masses. On the other hand, vector boson
fusion production of an invisibly decaying Higgs \cite{Sirunyan:2018owy} can already
significantly constraint the models parameter space for
$m_H\,\gtrsim\,62.5\,\GeV$. 
Depending
on the channel, masses of DM particle up to 200-300 GeV would be
accessible at the HL-LHC \cite{Datta:2016nfz,Dutta:2017lny,Wan:2018eaz}. For the di-jet plus missing transverse
energy signature, large background significantly affects LHC
sensitivity. 

Production cross sections for the benchmark points proposed here
  depend on the signature that is considered and can reach up to
  $\mathcal{O}\lb \pb \rb$ at the 13 \TeV LHC \cite{us_note,trtalk}. An
  increase in center-of-mass energy to 27 \TeV, as in the HE-LHC setup
  \cite{Zimmermann:2651305}, can lead to an increase of an order of
  magnitude in production cross sections. 
However, without detailed experimental analyses, projections of reachability are difficult to make. We therefore strongly encourage the experimental collaborations to investigate the benchmark points presented here at current and future LHC runs.}

\begin{center}
\begin{table}[ht!]
\small
\begin{center}
\begin{tabular}{|l|l|l|l|l|l|l|}
\hline
No. & $m_H$ & $m_A$ & $m_{H^\pm}$ & $\lambda_2$ & $\lambda_{345}$ & $\Omega_c h^2$\\\hline
\textbf{BP1} & 72.77 & 107.8 & 114.6 & 1.445 & -0.004407 & 0.1201 \\  \hline
BP2 & 65 & 71.53 & 112.8 & 0.7791 & 0.0004 & 0.07081 \\  \hline
BP3 & 67.07 & 73.22 & 96.73 & 0 & 0.00738 & 0.06162 \\  \hline
BP4 & 73.68 & 100.1 & 145.7 & 2.086 & -0.004407 & 0.08925 \\  \hline
BP6 & 72.14 & 109.5 & 154.8 & 0.01257 & -0.00234 & 0.1171 \\  \hline
BP7 & 76.55 & 134.6 & 174.4 & 1.948 & 0.0044 & 0.0314 \\  \hline
\textbf{BP8} & 70.91 & 148.7 & 175.9 & 0.4398 & 0.0058 & 0.122 \\  \hline
BP9 & 56.78 & 166.2 & 178.2 & 0.5027 & 0.00338 & 0.08127 \\  \hline
BP10 & 76.69 & 154.6 & 163 & 3.921 & 0.0096 & 0.02814 \\  \hline
BP11 & 98.88 & 155 & 155.4 & 1.181 & -0.0628 & 0.002737 \\  \hline
BP12 & 58.31 & 171.1 & 173 & 0.5404 & 0.00762 & 0.00641 \\  \hline
BP13 & 99.65 & 138.5 & 181.3 & 2.463 & 0.0532 & 0.001255 \\  \hline
\textbf{BP14} & 71.03 & 165.6 & 176 & 0.3393 & 0.00596 & 0.1184 \\ \hline 
\textbf{BP15} & 71.03 & 217.7 & 218.7 & 0.7665 & 0.00214 & 0.1222 \\  \hline
\textbf{BP16} & 71.33 & 203.8 & 229.1 & 1.03 & -0.00122 & 0.1221 \\  \hline
BP18 & 147 & 194.6 & 197.4 & 0.387 & -0.018 & 0.001772 \\  \hline
BP19 & 165.8 & 190.1 & 196 & 2.768 & -0.004 & 0.002841 \\  \hline
BP20 & 191.8 & 198.4 & 199.7 & 1.508 & 0.008 & 0.008494 \\  \hline
\textbf{BP21} & 57.48 & 288 & 299.5 & 0.9299 & 0.00192 & 0.1195 \\  \hline
\textbf{BP22} & 71.42 & 247.2 & 258.4 & 1.043 & -0.0032 & 0.122 \\  \hline
BP23 & 62.69 & 162.4 & 190.8 & 2.639 & 0.0056 & 0.06404 \\  \hline
\end{tabular}
\caption{ \label{tab:bench}
  Low mass IDM benchmark points considered in the presented study, taken from \cite{Kalinowski:2018ylg}.
  In all benchmarks $m_h = 125.1$ \GeV. Bold font denotes BP for
  which $H$ completely saturates DM relic density. Note that BP5 and BP17 
  were excluded by the updated {\sc Xenon1T} limits \cite{Aprile:2018dbl}.
The values of the $\lambda_{345}$ parameter for scenarios BP8 and BP22
were slightly modified with respect to \cite{Kalinowski:2018ylg}, to be
consistent with the most recent results on the relic density
\cite{Aghanim:2018eyx}.
}
\end{center}
\end{table}
\end{center}

\begin{center}
\begin{table}[ht !]
\small
\begin{center}
\begin{tabular}{|l|l|l|l|l|l|l|}
\hline 
 No. & $m_H$ & $m_A$ & $m_{H^\pm}$ & $\lambda_2$ & $\lambda_{345}$ & $\Omega_c h^2$\\ 
  \hline 
HP1 & 176 & 291.4 & 312 & 1.49 & -0.1035 & 0.0007216 \\ \hline 
HP2 & 557 & 562.3 & 565.4 & 4.045 & -0.1385 & 0.07209 \\ \hline 
HP3 & 560 & 616.3 & 633.5 & 3.38 & -0.0895 & 0.001129 \\ \hline 
HP4 & 571 & 676.5 & 682.5 & 1.98 & -0.471 & 0.0005635 \\ \hline 
HP5 & 671 & 688.1 & 688.4 & 1.377 & -0.1455 & 0.02447 \\ \hline 
HP6 & 713 & 716.4 & 723 & 2.88 & 0.2885 & 0.03515 \\ \hline 
HP7 & 807 & 813.4 & 818 & 3.667 & 0.299 & 0.03239 \\ \hline 
HP8 & 933 & 940 & 943.8 & 2.974 & -0.2435 & 0.09639 \\ \hline 
HP9 & 935 & 986.2 & 988 & 2.484 & -0.5795 & 0.002796 \\ \hline 
\textbf{HP10} & 990 & 992.4 & 998.1 & 3.334 & -0.040 & 0.122 \\ \hline 
HP11 & 250.5 & 265.5 & 287.2 & 3.908 & -0.1501 & 0.00535 \\ \hline 
HP12 & 286.1 & 294.6 & 332.5 & 3.292 & 0.1121 & 0.00277 \\ \hline 
HP13 & 336 & 353.3 & 360.6 & 2.488 & -0.1064 & 0.00937 \\ \hline 
HP14 & 326.6 & 331.9 & 381.8 & 0.02513 & -0.06267 & 0.00356 \\ \hline 
HP15 & 357.6 & 400 & 402.6 & 2.061 & -0.2375 & 0.00346 \\ \hline 
HP16 & 387.8 & 406.1 & 413.5 & 0.8168 & -0.2083 & 0.0116 \\ \hline 
HP17 & 430.9 & 433.2 & 440.6 & 3.003 & 0.08299 & 0.0327 \\ \hline 
HP18 & 428.2 & 454 & 459.7 & 3.87 & -0.2812 & 0.00858 \\ \hline 
HP19 & 467.9 & 488.6 & 492.3 & 4.122 & -0.252 & 0.0139 \\ \hline 
HP20 & 505.2 & 516.6 & 543.8 & 2.538 & -0.354 & 0.00887 \\ \hline 
\end{tabular}
\caption{ \label{tab:bench2}
  Additional set of high mass IDM benchmark points considered in the
  presented study, taken from \cite{Kalinowski:2018ylg}.
 In all benchmarks $m_h = 125.1$ \GeV. For HP10 scenario (bold)
  $H$ completely saturates DM relic density;
the value of the $\lambda_{345}$ parameter for this scenario
was slightly modified with respect to \cite{Kalinowski:2018ylg}, to be
consistent with the most recent results on the relic density \cite{Aghanim:2018eyx}.
}
\end{center}
\end{table}
\end{center}




\section{Analyses strategies and simulation setup}
\label{Simulation}

In this work, we consider the following tree-level production processes of inert scalars
at $e^+ e^-$ collisions\footnote{The process $e^+e^-\to AA$ is in
  principle possible as well. However, this process is
  moderated by an $s-$channel SM Higgs and therefore highly suppressed by
  the electron Yukawa couplings.} 
\bea\label{eq:procs}
 e^+ e^- & \to &  A \; H,  \\
 e^+ e^- & \to &  H^+ H^-. \nonumber 
\eea
For the calculation of cross-sections as well as detailed signal and
background simulation, we make use of the Monte Carlo event generator {\tt WHizard 2.2.8}
\cite{Moretti:2001zz,Kilian:2007gr}. For the signal, we employ the IDM model implemented in 
SARAH \cite{Staub:2015kfa}. Model parameter files for the considered benchmark
scenarios were prepared using {\tt SPheno 4.0.3} \cite{Porod:2003um,Porod:2011nf}.
When generating signal and background events samples for the presented
analysis, energy spectra for CLIC \cite{Linssen:2012hp}, based on
detailed beam simulations, were taken into account.
{For initial state radiation (ISR), the intrinsic {\tt WHizard}
implementation of the lepton ISR structure function includes all
orders of soft and soft-collinear photons as well as up to the third
order in hard-collinear photons.}

\begin{figure}[ht]
\begin{center}
  \includegraphics[width=0.48\textwidth]{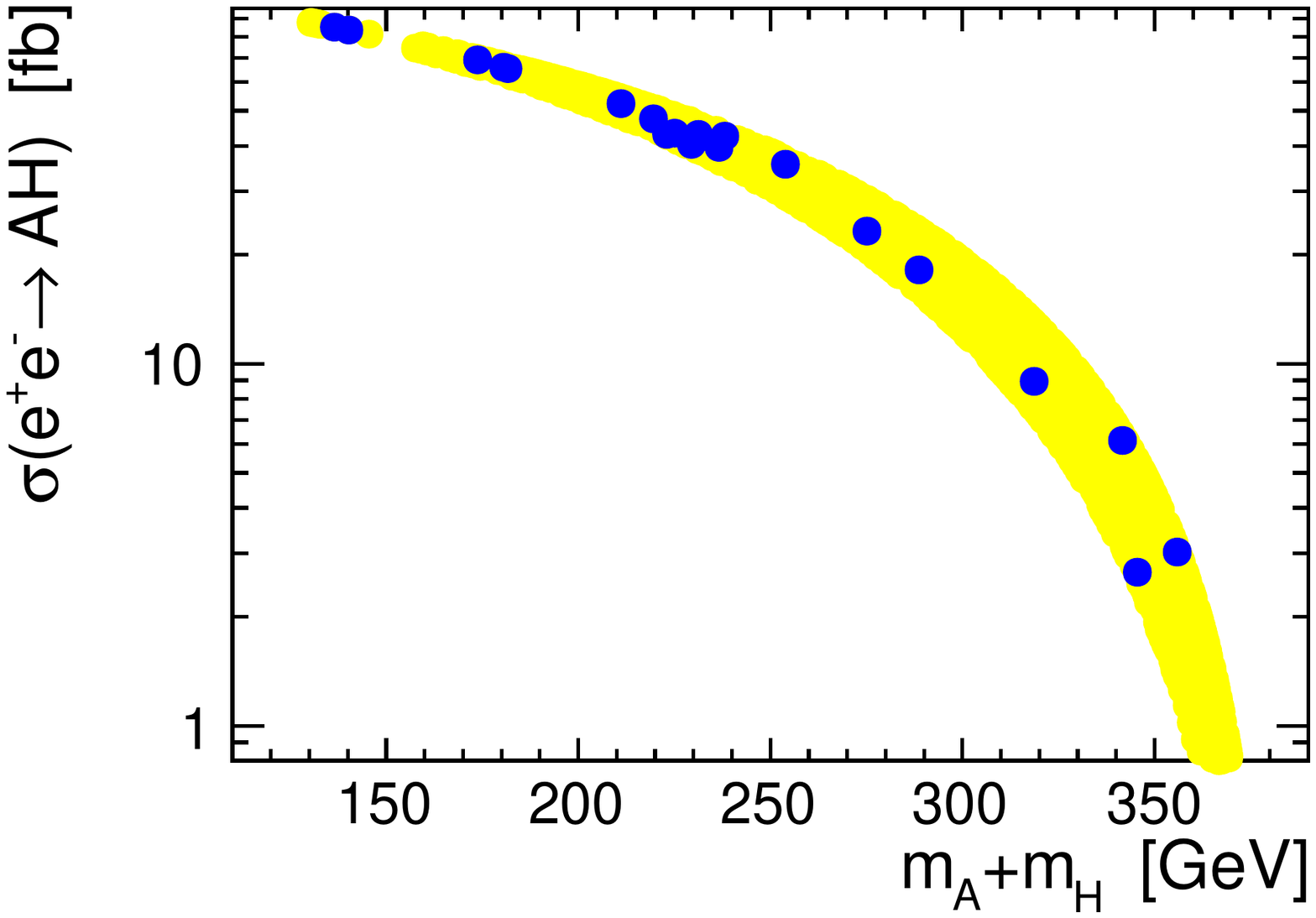}
  \includegraphics[width=0.48\textwidth]{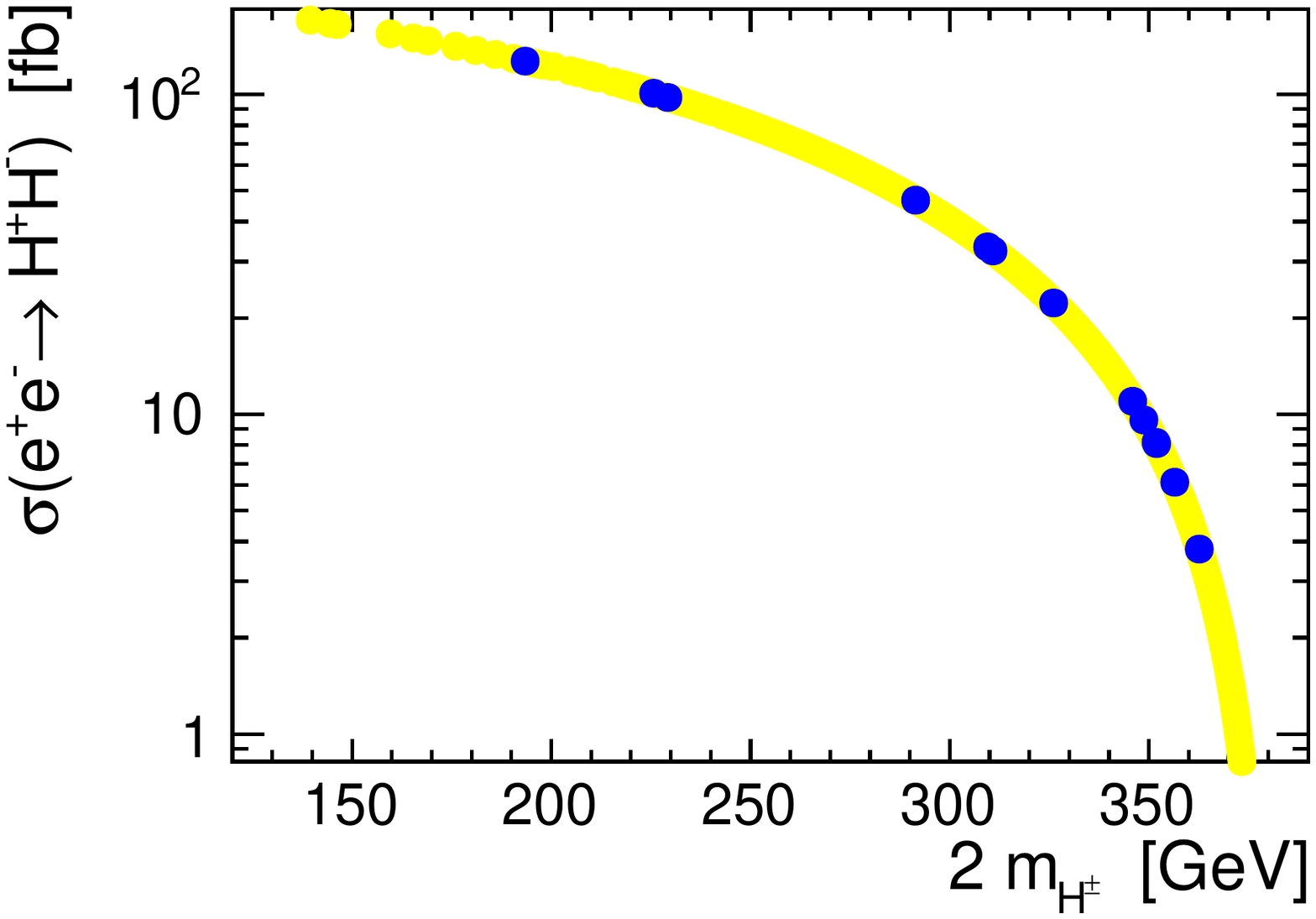}
\end{center}  
\caption{Leading-order cross-sections for neutral {\sl (left)} and charged
  {\sl (right)} inert scalar production, $e^+e^-\to H A$ and
   $e^+e^-\to H^+ H^-$, for 380\,\GeV~collision energy. The yellow band
  represents all scenarios selected in the model scan \cite{Kalinowski:2018ylg} while
  the blue dots represent the selected benchmark scenarios. Beam energy
  spectra are not included. 
}
\label{fig_ahcros}
\end{figure}

Leading-order cross-sections for the processes in (\ref{eq:procs}) for
380\,\GeV~collision energy, including initial state radiation, are
presented in \cref{fig_ahcros}. In the scenarios considered in this paper the produced dark scalar $A$
decays to a (real or virtual) $Z$ boson and the (lighter)
neutral scalar $H$, $A \rightarrow Z^{(\star)}H$, while the produced
charged boson $H^\pm$ decays predominantly to a (real or virtual) $W^\pm$ boson
and the neutral scalar $H$, $H^+ \rightarrow {W^\pm}^{(\star)}H$,
{where the DM candidate $H$ escapes detection}. 
Since both the production and decay processes are governed by the SM electroweak couplings,  the
inert masses are the only BSM parameters probed at CLIC.

Since isolated leptons (electrons and muons) can be identified and
reconstructed with very high efficiency and
{purity}~\cite{Linssen:2012hp}, we concentrate on $Z$ and $W^\pm$
leptonic decays, leading to a signature of leptons and missing
transverse energy. 
{For the same flavour lepton pair signature, we restrict the
  analysis to the $\mu^+\mu^-$ final state, allowing for almost perfect
  reconstruction of lepton kinematics\footnote{{For final
    state electrons energy resolution is affected by the
    final state radiation and bremsstrahlung
    effects~\cite{Linssen:2012hp}.}},  while for different flavour 
  lepton pairs $\mu^+ e^-$ and $e^+ \mu^-$ final states are considered.} 
We refrain from including detector effects in the results presented
here as {for the considered final states} they are expected to 
only marginally change the outcome of our study.
Muon pair production can be a signature of the $AH$ production process
followed by the $A$ decay:
\bea
e^+ e^- & \to & H A \; \to \;  H H Z^{(\star)} \; \to \;   H H \mu^+\mu^- 
\label{neutral}
\eea
while the production of the different flavour lepton pair is the
expected signature for $H^+H^-$ production:
\bea
e^+ e^- & \to &  H^+ H^- \; \to \; H H {W^+}^{(*)} {W^-}^{(*)} \; \to \;  H H \lep^+ \lep'^-  \nu\bar\nu'.
\label{charged}
\eea
Note that when both $W$ bosons in (\ref{charged}) decay to muons, 
the charged Higgs pair production process will contribute to the  signature (\ref{neutral}) of $HA$ production as well. 

During simulations we do not constrain the intermediate particles, but
consider all processes leading to
$\ell^+\,(\ell^-)'\,+\,\slashed{E}_\perp$. Especially processes with
additional neutrinos can contribute and need to be taken into
account. This includes processes with tau (pair) production and
their successive leptonic decays.

To be specific, for processes with muons in the final state, the
following processes have been simulated:
\bea \label{eq:neutralsignal}
e^+ e^- & \to &  \mu^+ \mu^-  \; H H, \nonumber  \\
        & \to &  \mu^+ \mu^-  \nu_\mu \bar{\nu}_\mu \; H H, \nonumber \\
        & \to &  \tau^+ \mu^-  \nu_\tau \bar{\nu}_\mu \; H H, \;\;
                 \mu^+ \tau^-  \nu_\mu \bar{\nu}_\tau \; H H, \nonumber \\
        & \to &  \tau^+ \tau^-  \; H H, \nonumber \;\;
                 \tau^+ \tau^-  \nu_\tau \bar{\nu}_\tau \; H H,
\eea
where the final state taus are then forced to decay to a muon and a neutrino. For the background the following Standard Model processes are considered:
\bea
e^+ e^- & \to &  \mu^+ \mu^-,\nonumber  \\
        & \to &  \mu^+ \mu^-  \; \nu_i \bar{\nu}_i, \nonumber \\
        & \to &  \tau^+ \mu^-  \; \nu_\tau \bar{\nu}_\mu, \; \mu^+ \tau^-  \; \nu_\mu \bar{\nu}_\tau, \nonumber\\
        & \to &  \tau^+ \tau^-, \;   \tau^+ \tau^-  \; \nu_i \bar{\nu}_i, \nonumber
\eea
where the additional neutrino pair can be of any flavour ($i  = e, \mu, \tau$). As before, we generate all 
processes leading to the above final states, {without} constraining the intermediate particles states.

Similarly, for the electron-muon pair final state the following signal
processes are considered:
\bea \label{eq:chargedsignal}
e^+ e^- & \to &  \mu^+ \nu_\mu \; e^-  \bar{\nu}_e \; H H, \;\;
                 e^+ \nu_e \; \mu^-  \bar{\nu}_\mu \; H H, \nonumber \\
        & \to &  \mu^+ \nu_\mu \; \tau^-  \bar{\nu}_\tau \; H H, \;\;
                 \tau^+ \nu_\tau \; \mu^-  \bar{\nu}_\mu \; H H, \nonumber \\
        & \to &  e^+ \nu_e \; \tau^-  \bar{\nu}_\tau \; H H, \;\;
                 \tau^+ \nu_\tau \; e^-  \bar{\nu}_e \; H H, \nonumber \\
        & \to &  \tau^+ \; \tau^- \; H H, \nonumber \;\;
                 \tau^+ \nu_\tau \; \tau^-  \bar{\nu}_\tau \; H H, 
\eea
with the final state tau leptons decaying to an electron or a muon (to
match the required final state signature, $\mu^+ e^-$ or $e^+\mu^-$).
For the background in this case the following Standard Model four-fermion processes are considered:
\bea \label{eq:chargedbg}
e^+ e^- & \to &  \mu^+ \nu_\mu \; e^-  \bar{\nu}_e \; , \;  e^+ \nu_e \; \mu^-  \bar{\nu}_\mu \; , \nonumber \\
        & \to &  \mu^+ \nu_\mu \; \tau^-  \bar{\nu}_\tau \; , \; \tau^+ \nu_\tau \; \mu^-  \bar{\nu}_\mu , \nonumber \\
        & \to &  e^+ \nu_e \; \tau^-  \bar{\nu}_\tau \; , \;  \tau^+ \nu_\tau \; e^-  \bar{\nu}_e \; , \nonumber \\
        & \to & \tau^+ \tau^-, \;    \tau^+ \nu_\tau \; \tau^-  \bar{\nu}_\tau \;.\nonumber
\eea

As {mentioned} above, beam energy spectra and ISR were taken into account.

Generator-level cuts corresponding to the expected detector acceptance
were applied for both signal and background simulations: electrons or
muons with energy of at least 5~\GeV~need to be emitted at least
100~mrad from the beam direction and their angular 
separation should also be at least 100~mrad. To reduce background from radiative Z-return events we also require
that there are no ISR photons emitted at angles above 100~mrad with
energies larger that 10~\GeV.

For the considered final states we assume that only two charged
leptons are reconstructed in the detector.
The observed final state can be completely described by a small set of
kinematic variables. To assure the best possible discrimination between signal and
background events, resulting in highest expected significance of the
possible observation, we make use of multivariate analyses.
We apply the Boosted Decision Tree (BDT) classification algorithm, as implemented in TMVA
toolkit \cite{Hocker:2007ht}, with the following eight input variables describing the kinematics of the dilepton final state:
\begin{itemize}
\item total energy of the lepton pair, E$_{\ell\ell}$;
\item dilepton invariant mass, $M_{\ell\ell}$;
\item dilepton transverse momentum, p$_\text{T}^{\ell \ell}$;
\item polar angle of the dilepton pair, $\Theta_{\ell\ell}$ ;
\item Lorentz boost of the dilepton pair, $\beta_{\ell\ell} = \text{p}_{\ell\ell}/\text{E}_{\ell\ell}$;
\item reconstructed missing (recoil) mass $M_\text{miss}$ (calculated
  assuming   nominal $e^+e^-$ collision energy);
\item $\ell^{-}$ production angle with respect to the beam direction,
  calculated in the dilepton center-of-mass frame, $\Theta^{\star}_{\ell}$; 
\item $\ell^{-}$ production angle with respect to the dilepton pair
  boost direction, calculated in the dilepton center-of-mass
  frame, $\angle^{\star}(\ell,\ell\ell)$,
\end{itemize}
where lepton pair $\ell\ell$ denotes $\mu^+ \mu^-$ for $AH$ channel and
$\mu^+ e^-$ or $e^+ \mu^-$ for $H^+H^-$ production.
The first five variables refer to the dilepton pair system as a whole, while the last two correspond to the single lepton polar angle calculated in the two different reference frames.
Please note that these eight variables are not independent, as the final state with two massless leptons and missing energy only can be completely described by five parameters (plus azimuthal angle, which is not relevant).
However, using more input variables resulted in better signal selection efficiencies.
The BDT algorithm is trained individually for each benchmark scenario
and each running energy using the generated event samples after
detector acceptance and pre-selection cuts.



\section{Inert Scalars at the first stage of CLIC}
\label{Low energy}

First we investigate the discovery prospects for the IDM benchmarks at
the initial CLIC operation at $\sqrt{s}=380$\GeV with an expected integrated
luminosity of 1\abinv \cite{Robson:2018zje}.

The possibility to access benchmark points with $\sum
m_i\,\geq\,380\GeV$, that are not accessible at the first stage, is
investigated in section \ref{High energy}, where 
the second and third energy stages of CLIC, at 1.5\,TeV and 3\,TeV are considered 
with integrated luminosities of 2.5 and 5\abinv, respectively.

\subsection{Neutral dark scalar pair production $e^{+}e^{-}\rightarrow AH$}

\begin{figure}[tb]
\begin{center}
  \includegraphics[width=0.6\textwidth]{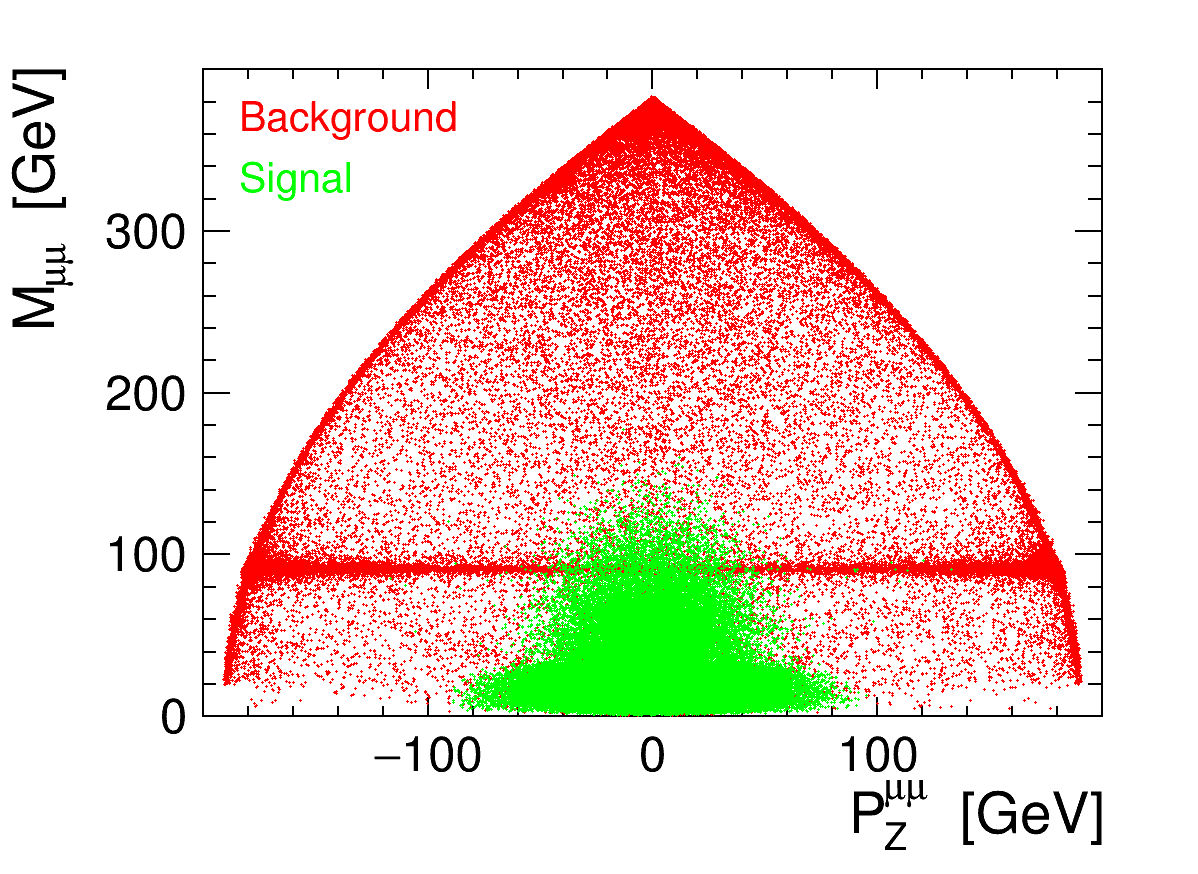}
\end{center}  
\caption{Distribution of the lepton pair invariant mass, M$_{\mu\mu}$,
  as a function of the lepton pair longitudinal momentum,
  P$_\text{Z}^{\mu\mu}$, for BP1 scenario (green points) and Standard Model
  background (red points). Same number of events were simulated for signal
  and background for centre-of-mass energy of 380\GeV, using CLIC
  luminosity spectra.}
\label{fig_preselAH}
\end{figure}

As described above, in this channel we focus on final
states with muon  pairs and missing transverse energy. As the DM
particles escape detection, the signal process will lead to large
missing energy and momentum. Furthermore, the invariant mass of the
lepton pair, stemming 
from the decay of a real or virtual $Z$ boson, should be relatively small
(depending on the mass splitting between $A$ and $H$, but not greater than
$m_{Z}$). 
On the other hand, the dominant Standard Model background process {proceeds via the} 
$s$-channel $Z/\gamma$ di-muon production, with most pairs produced either with high invariant mass (events
without hard ISR) or with significant longitudinal boost (events with
high-energy ISR photon).
We display the lepton pair invariant mass distribution for signal and
background processes in \cref{fig_preselAH}. As expected, we observe that the event distribution for the signal 
(for the benchmark
scenario BP1; green points) is concentrated on a much smaller range in
the $P_z^{\mu\mu}, M^{\mu\mu}$ plane than the SM background distribution (red points).
For the 380\GeV analysis we therefore require an invariant mass
of the produced lepton pair to be below 100\GeV, and the absolute
value of the longitudinal momentum below 140\GeV. 
These pre-selection cuts significantly reduce background from
direct two fermion production, $e^+ e^- \to \mu^+ \mu^-$, hardly
affecting the signal.

\begin{figure}[tb]
\begin{center}
  \includegraphics[width=0.49\textwidth]{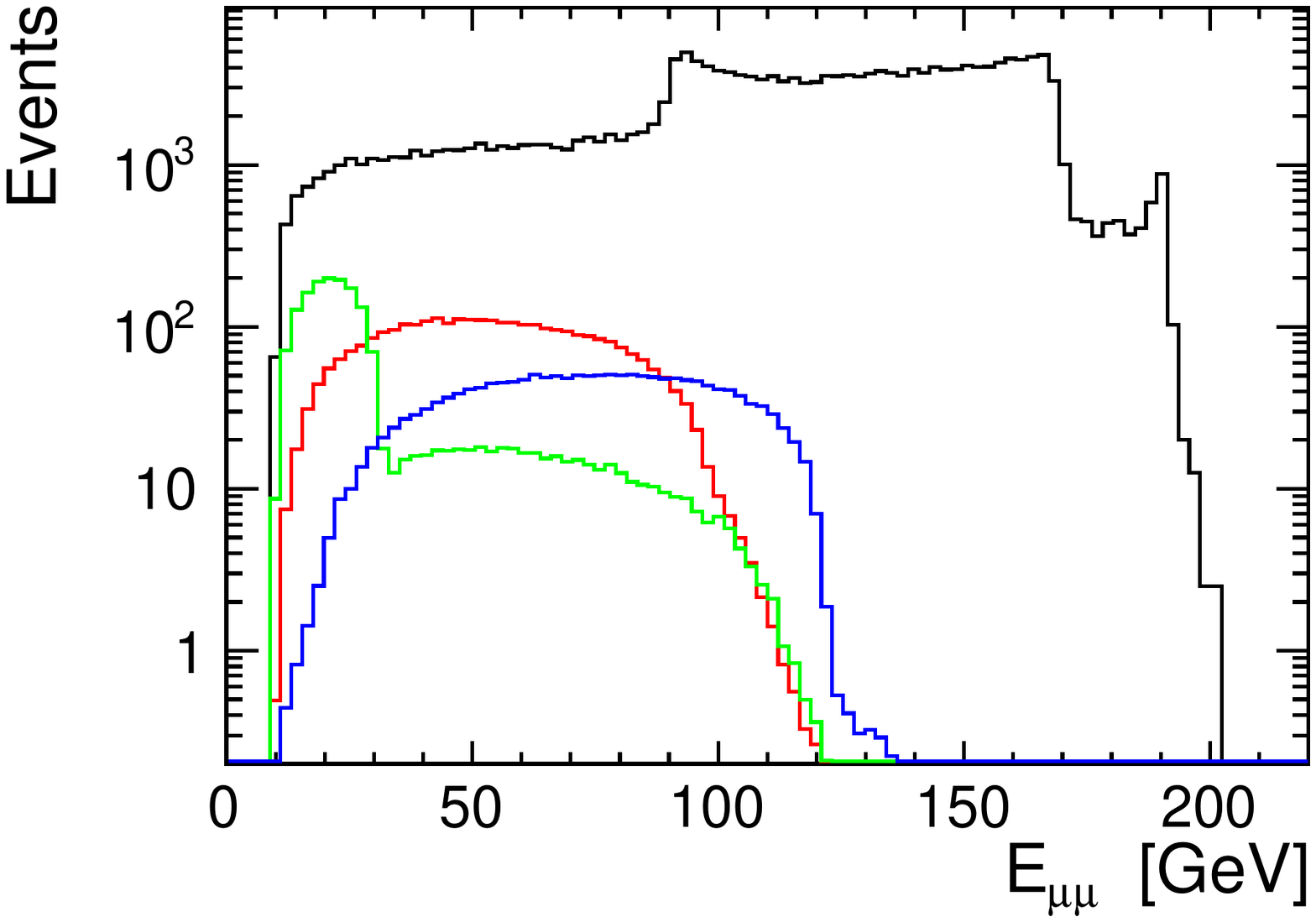}
  \includegraphics[width=0.49\textwidth]{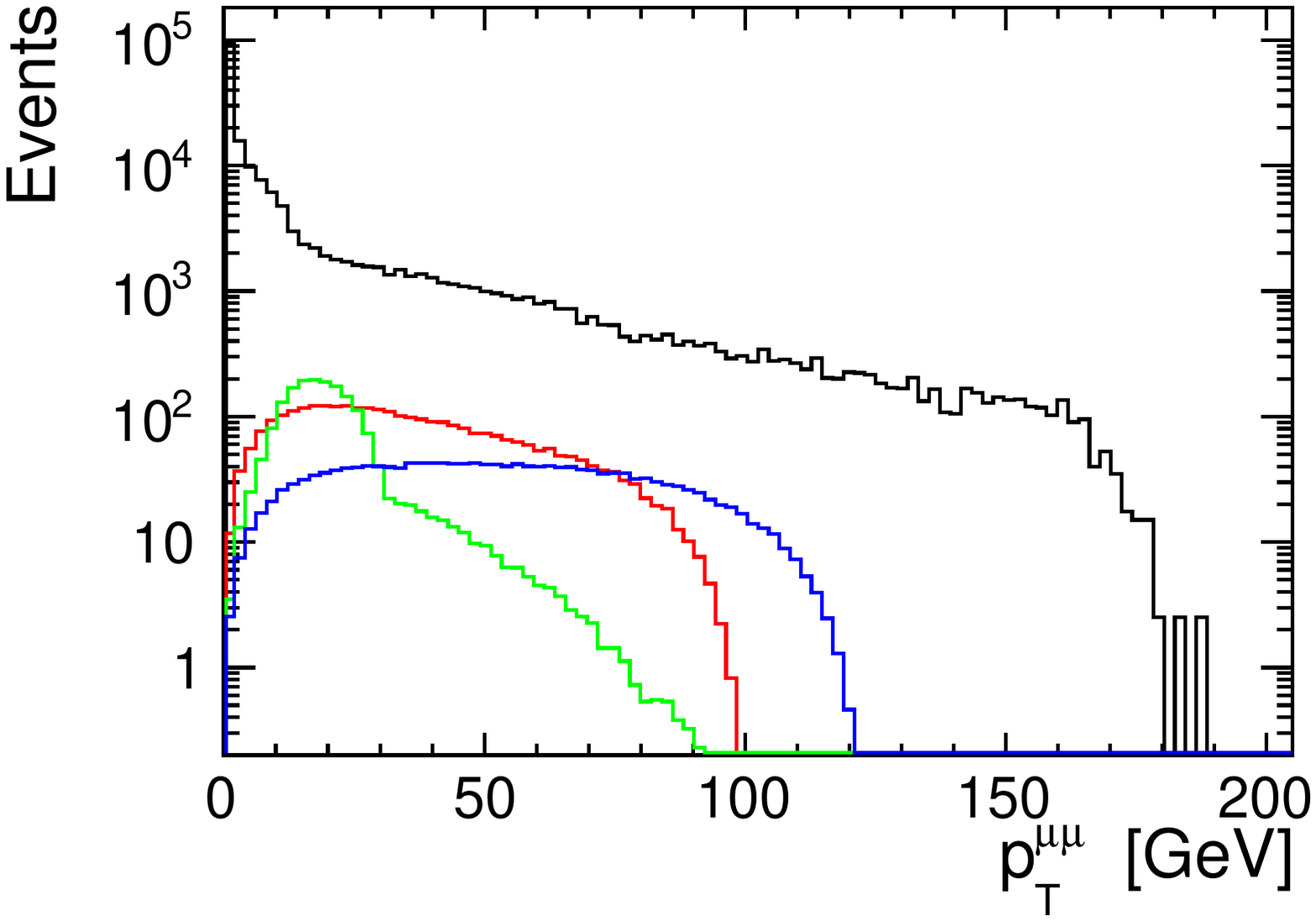}
  \includegraphics[width=0.49\textwidth]{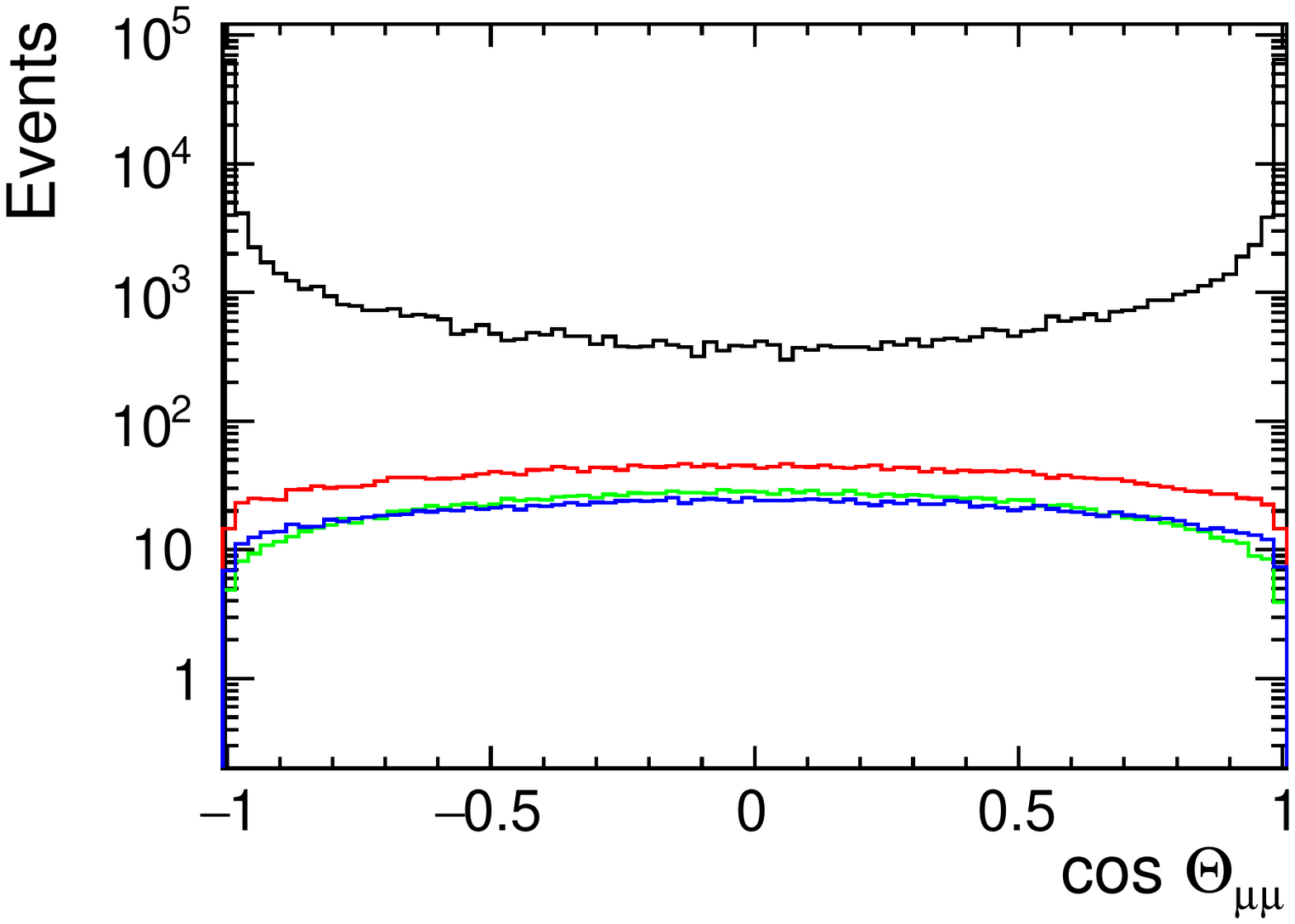}
  \includegraphics[width=0.49\textwidth]{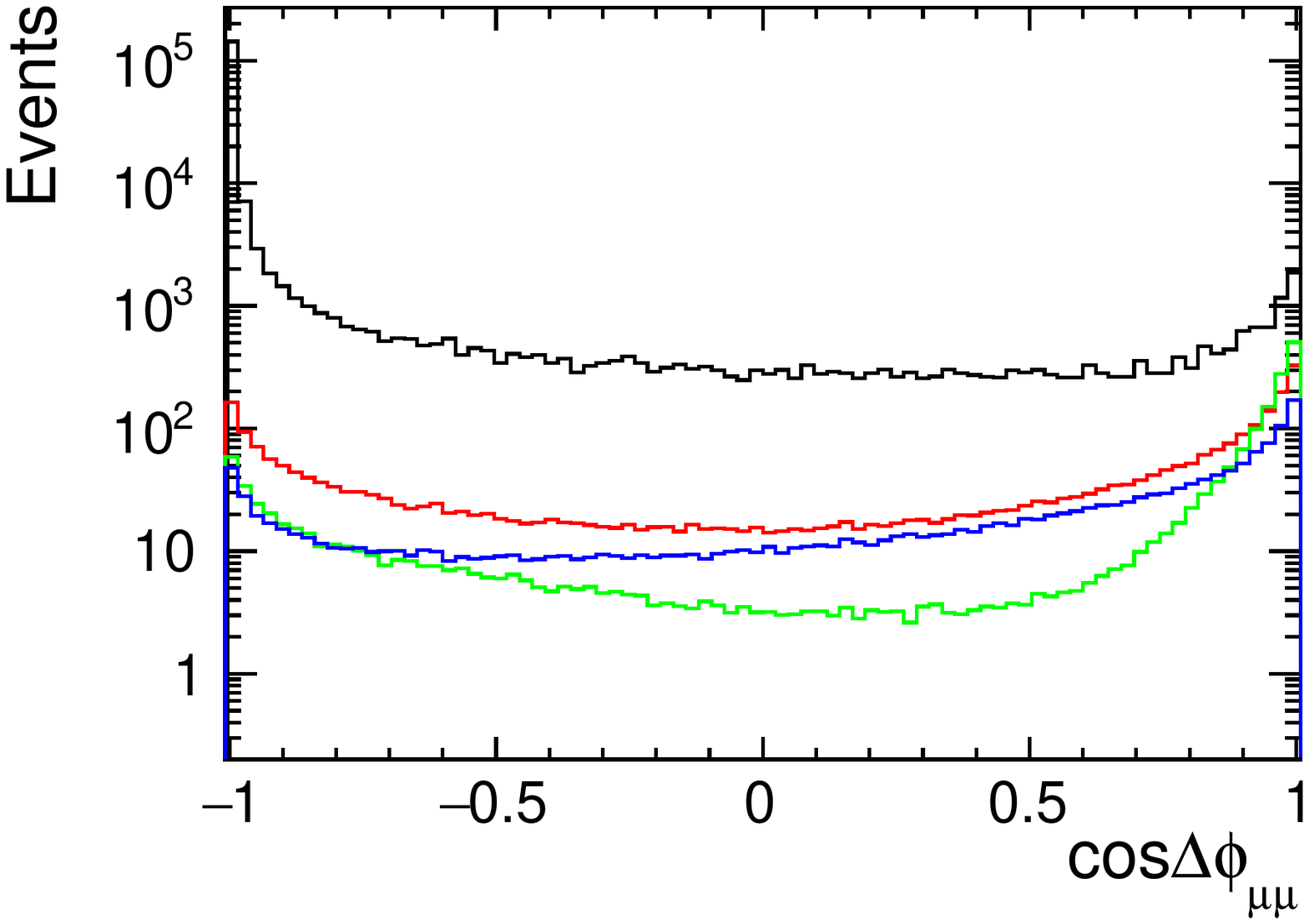}
\end{center}  
\caption{Distributions of the kinematic variables describing the
  leptonic final state considered in $AH$ analysis: lepton pair
energy, E$_{\mu\mu}$, total transverse momentum, p$^{\mu\mu}_\text{T}$,
pair production angle, $\cos\Theta_{\mu\mu}$ and the difference
of the lepton azimuthal angles, $\cos\Delta\phi_{\mu\mu}$.
Expected distributions for representative benchmarks BP1 (red histogram), BP2 (green) and BP7
(blue) are compared with expected background (black
histogram). Samples simulated for CLIC running at 380\GeV are
normalised to 1\abinv.} 
\label{fig_plotsAH}
\end{figure}

Shown in \cref{fig_plotsAH} are the distributions of the muon pair
energy, E$_{\mu\mu}$, total transverse momentum, p$^{\mu\mu}_\text{T}$,
polar angle, $\Theta_{\mu\mu}$, and the difference of the lepton
azimuthal angles, $\cos\Delta\phi_{\mu\mu}$ for three benchmark
scenarios {with virtual $Z$ boson production} and the SM background.
{As the experimental signature is expected to 
depend on the mass difference between $A$ and $H$ states, we present
benchmark points corresponding to different mass splittings:
of about 6\GeV (BP2), 35\GeV (BP1) and 58\GeV (BP7).}
For low mass differences, the contribution from $H^+H^-$ channel {to the muon pair production signature} is also
clearly visible (a tail of events with higher lepton pair energy and
transverse momentum).

\begin{figure}[tb]
\begin{center}
  \includegraphics[width=0.6\textwidth]{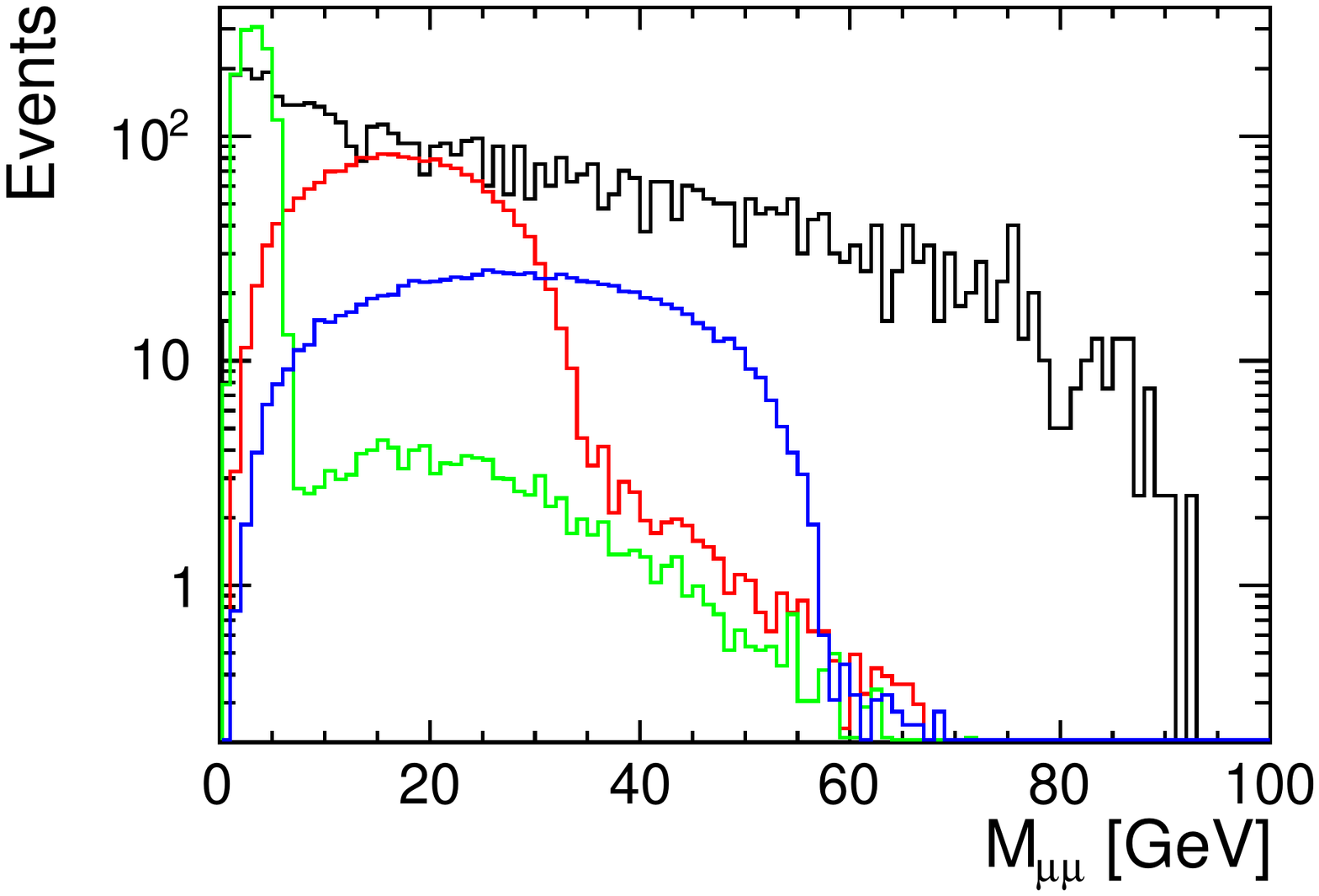}
\end{center}  
\caption{Distribution of the lepton pair invariant mass, M$_{\mu\mu}$,
  for BP1 (red histogram), BP2 (green) and BP7 (blue) signal
  scenarios, compared with expected Standard Model background (black
  histogram), after final event selection cuts (see text for details). 
  Samples simulated for CLIC running at 380\GeV are
  normalised to 1\abinv. }
\label{fig_cutsAH}
\end{figure}

Distributions of variables presented in \cref{fig_plotsAH} can be used
to select signal-enhanced samples of events.
The following selection requirements are therefore further imposed {on the lepton pair}:
\begin{itemize}
\item energy E$_{\mu\mu} < 100$\GeV,
\item transverse momentum p$^{\mu\mu}_\text{T} > 10$\GeV,
\item production angle (polar angle of the Z boson)
  $30^\circ < \Theta_{\mu\mu} < 150^\circ$,
\item difference of the lepton azimuthal angles $|\Delta\varphi_{\mu\mu}|<\frac{\pi}{2}$.
\end{itemize}
Presented in \cref{fig_cutsAH} is the lepton pair invariant mass
distribution after pre-selection and selection cuts.
Signal samples for selected benchmark scenario and the background
sample are normalised to 1\abinv.
About 5400 background events are expected after all selection cuts,
while 1810, 1290 and 540 signal events are expected for the BP1, BP2 and
BP7 scenarios, respectively.
This corresponds to about 21$\sigma$, 16$\sigma$ and $7\sigma$
significance.

\begin{figure}[tb]
\begin{center}
  \includegraphics[width=0.6\textwidth]{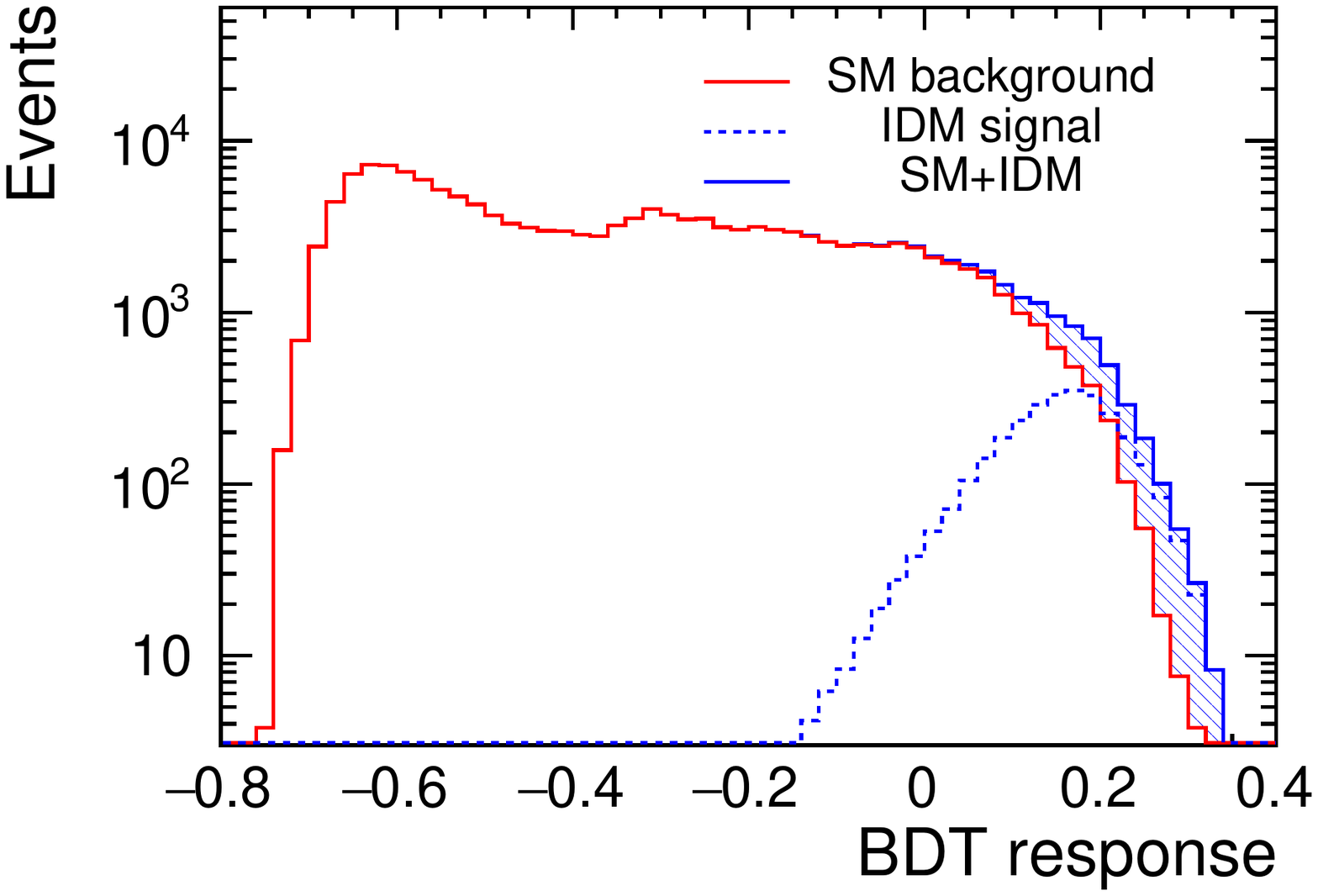}
\end{center}  
\caption{Response distributions of the BDT classifiers used for the
    selection of $AH$ production events at CLIC, at $\sqrt{s}=380$\GeV.
    Signal samples for BP1 scenario and SM background are normalised to
    1\abinv.}
\label{fig_bdtAH}
\end{figure}

\begin{figure}[tb]
\begin{center}
  \includegraphics[height=0.35\textwidth]{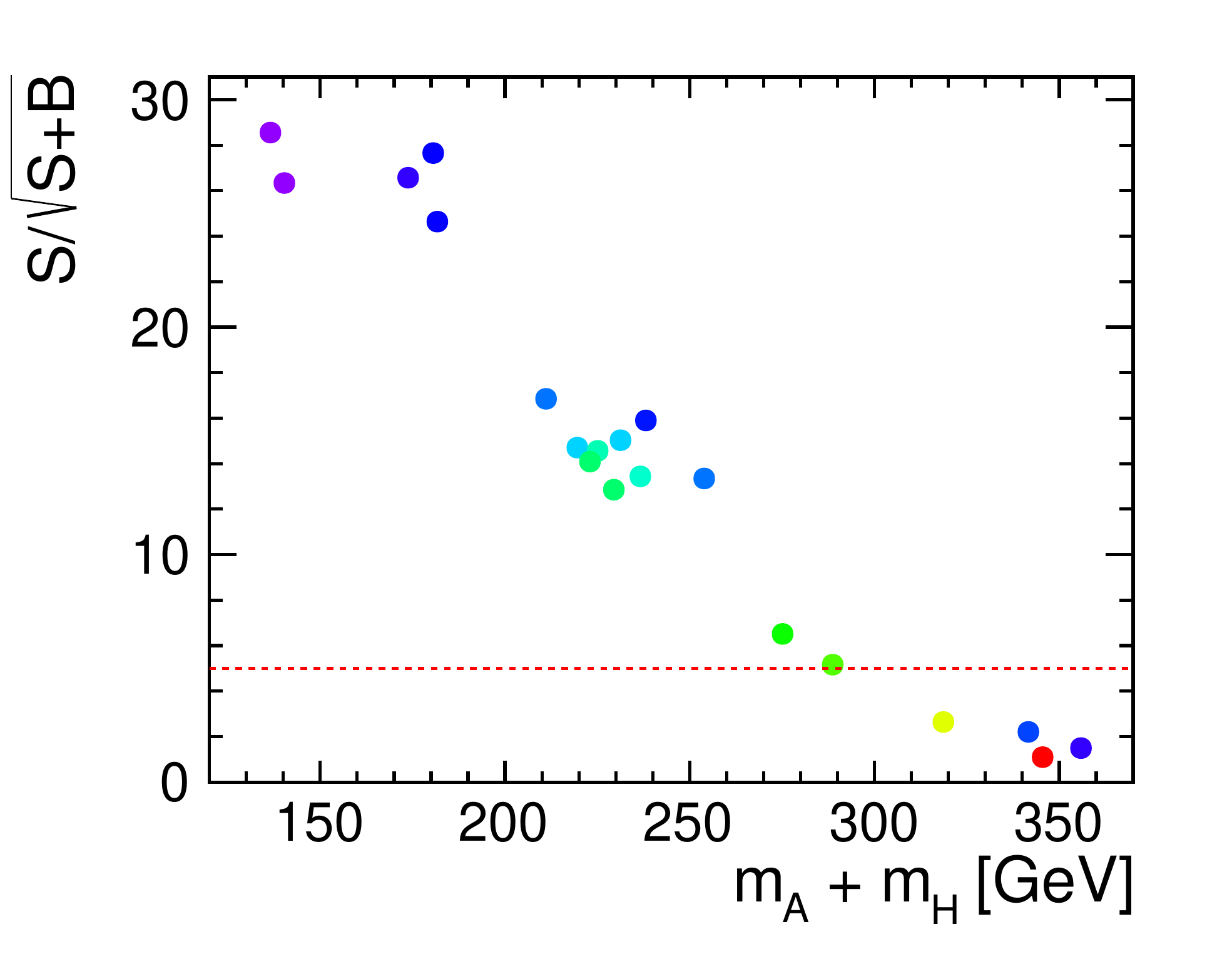}
  \includegraphics[height=0.35\textwidth]{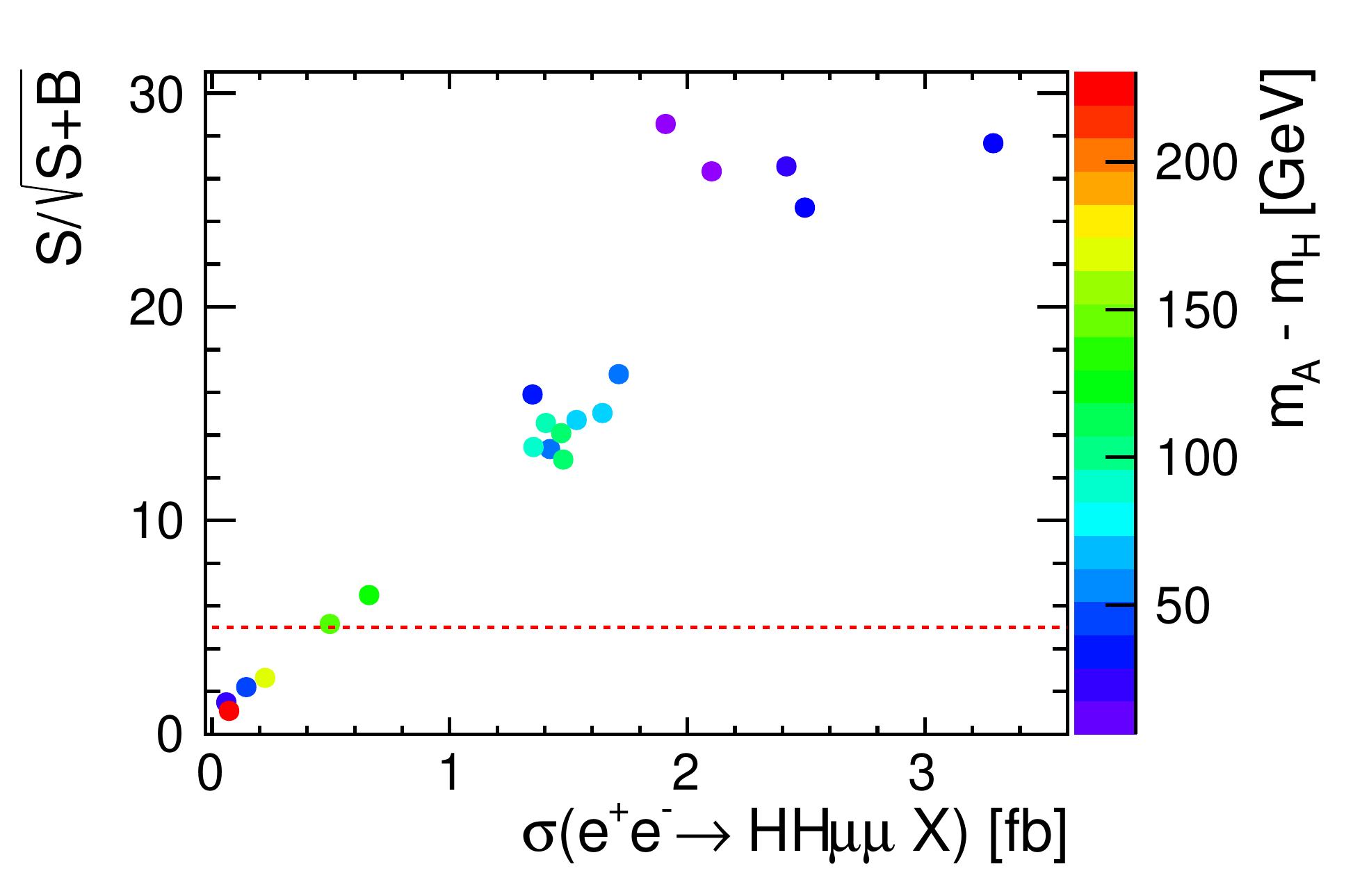}
\end{center}  
\caption{Expected significance of the deviations from the
  Standard Model predictions observed at 380\GeV CLIC for events with
  two muons in the final state ($\mu^+\mu^-$) as a function of the
  neutral inert scalar mass sum {\sl (left)} and the production cross-section 
  for the considered signal channel, after pre-selection cuts
  {\sl (right)}, for the BPs from table \ref{tab:bench}. 
  Color indicates the mass splitting between the A and H scalars 
  (right scale applies to both plots).
  } 
\label{fig_sigAH}
\end{figure}

Higher signal significances are obtained making use of 
multivariate analyses after the application of 
pre-selection cuts. 
As an example, we show the BDT response distributions for BP1 
(signal and SM background) in \cref{fig_bdtAH} for 1\abinv
collected at CLIC 380\GeV center-of-mass energy.
The optimal significance is obtained for a
BDT response cut of about 0.12, corresponding to 71\% signal selection
efficiency and 2.2\% background selection efficiency, with a resulting signal significance of about 27.7$\sigma$.
In \cref{fig_sigAH} the significance using the above method is displayed as a function
of the neutral inert scalar mass sum, $m_A + m_H$, and of the signal
production cross-section for the considered final state,
$\sigma(e^+e^-\rightarrow HH\mu^+\mu^- X_\text{inv})$.
The expected significance is mainly related to the $AH$ production cross-section. 
A discovery, corresponding to $5\,\sigma$, at the initial stage of
CLIC is expected for scenarios with the signal cross-section (in the
$\mu^+\mu^-$ channel, after pre-selection cuts on generator level)
above about 0.5\fb, which corresponds to the neutral inert scalar mass
sum below about 290\GeV.  
For the considered benchmark points we do not observe any sizable dependence 
of the expected significance on the mass splitting between the two neutral scalars,
$m_{A} - m_{H}$ (indicated by colour scale in \cref{fig_sigAH}).

\subsection{Charged scalar pair production $e^+ e^- \rightarrow H^+ H^-$}

The selection of $H^+H^-$ production events is more challenging than for
the $AH$ channel, as the two leptons in the final state no longer
originate from a single (on- or off-shell) intermediate state.
We therefore do not apply any additional pre-selection cuts  (except for the detector acceptance cuts, as described
in \cref{Simulation}). However, this also allows us to consider the
electron-muon pairs in the final state, avoiding large SM background from
the direct lepton pair production ($e^+e^- \rightarrow \ell^+ \ell^-$; this
channel contributes only via leptonic tau decays, suppressed by the
corresponding branching fractions).

\begin{figure}[tb]
\begin{center}
  \includegraphics[width=0.6\textwidth]{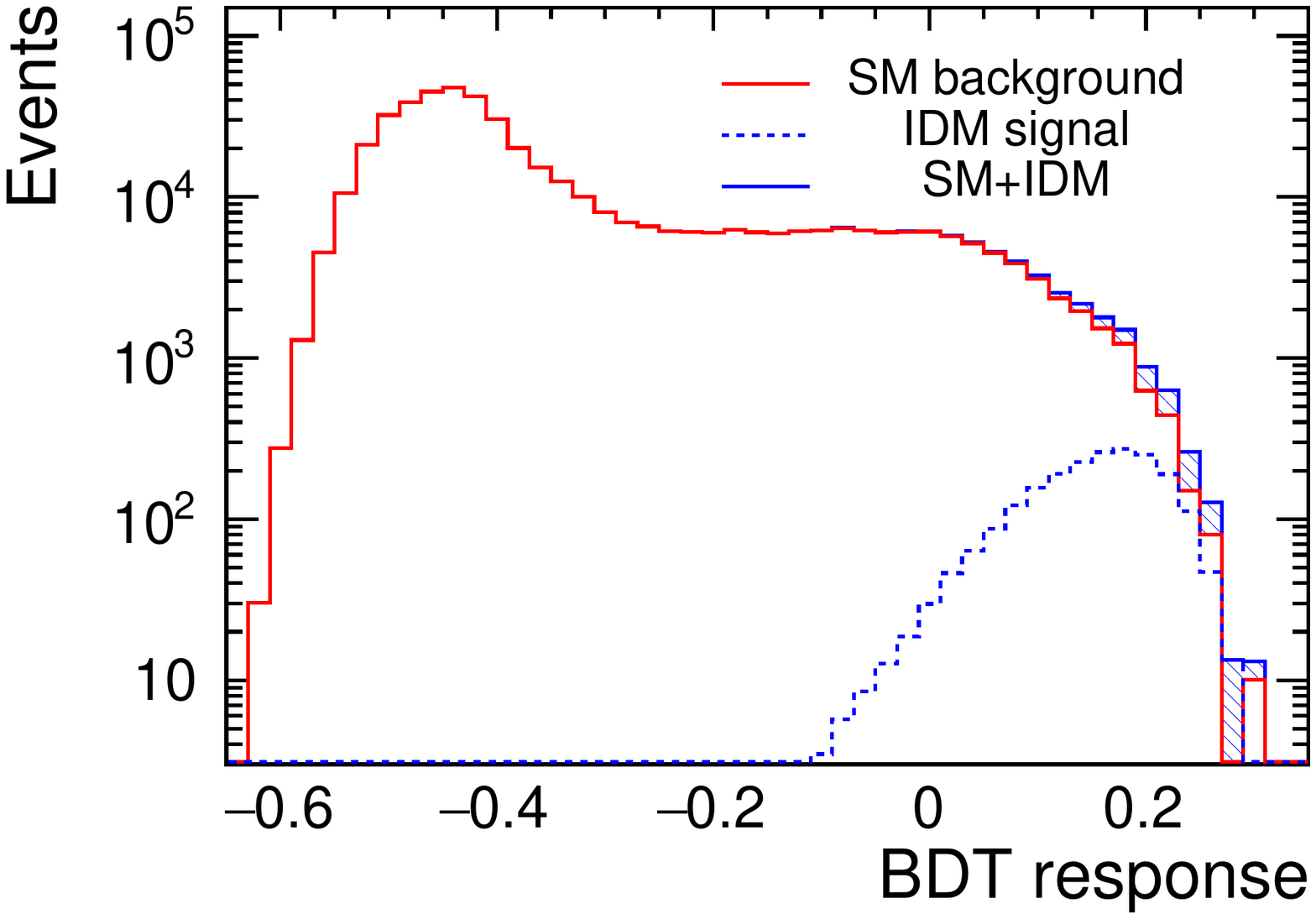}
\end{center}  
\caption{Response distributions of the BDT classifiers used for the
    selection of $H^+H^-$ production events at CLIC, at $\sqrt{s}=380\GeV$.
    Signal sample for BP1 scenario and SM background are normalised to
    1\abinv.}
\label{fig_bdtHpHm}
\end{figure}

With only the detector acceptance cuts on the generator level, the expected
background cross-section for the considered final state is about
500\fb,  over two orders of magnitude higher than for the
considered benchmark points.
\begin{figure}[tb]
\begin{center}
  \includegraphics[width=0.49\textwidth]{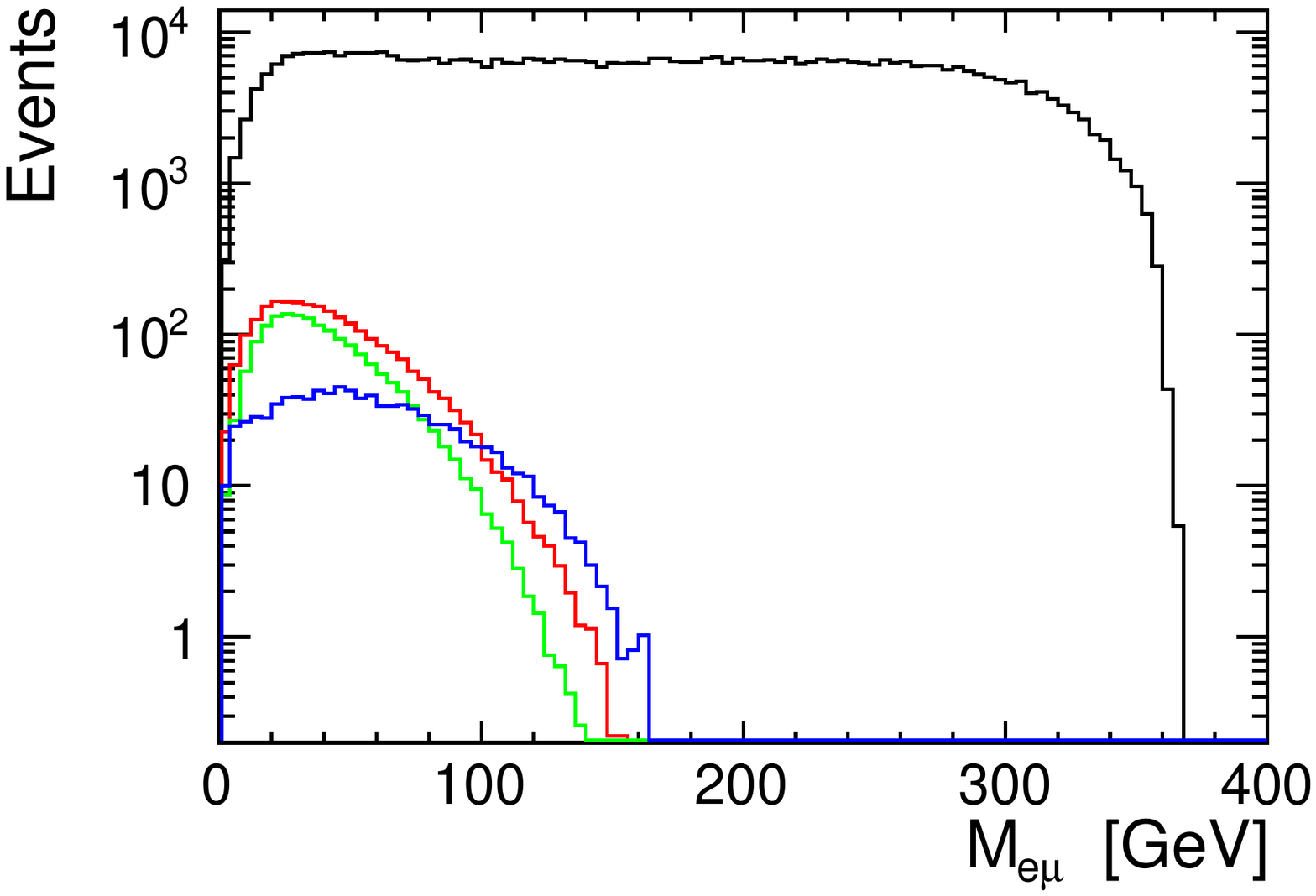}
  \includegraphics[width=0.49\textwidth]{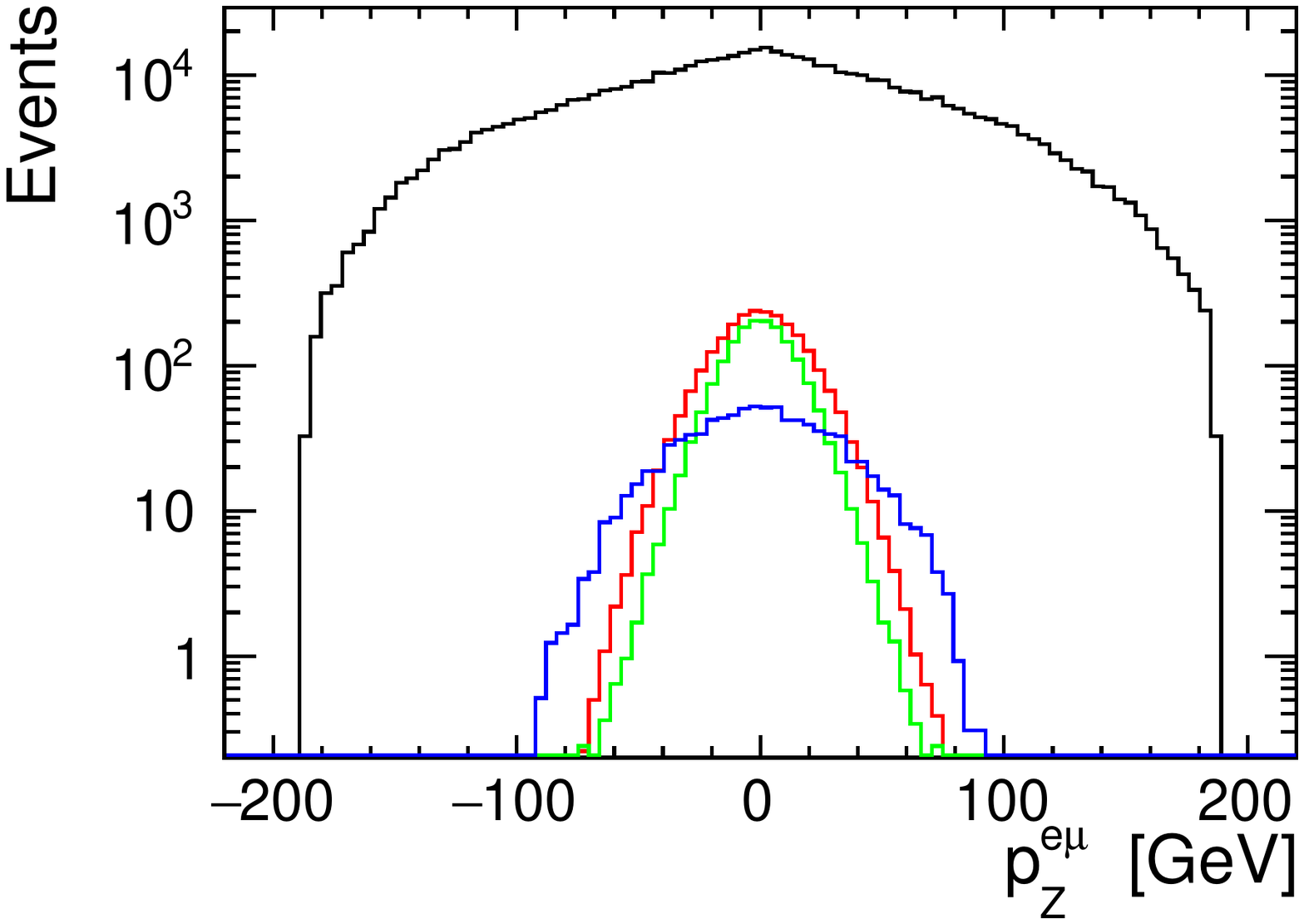}
  \includegraphics[width=0.49\textwidth]{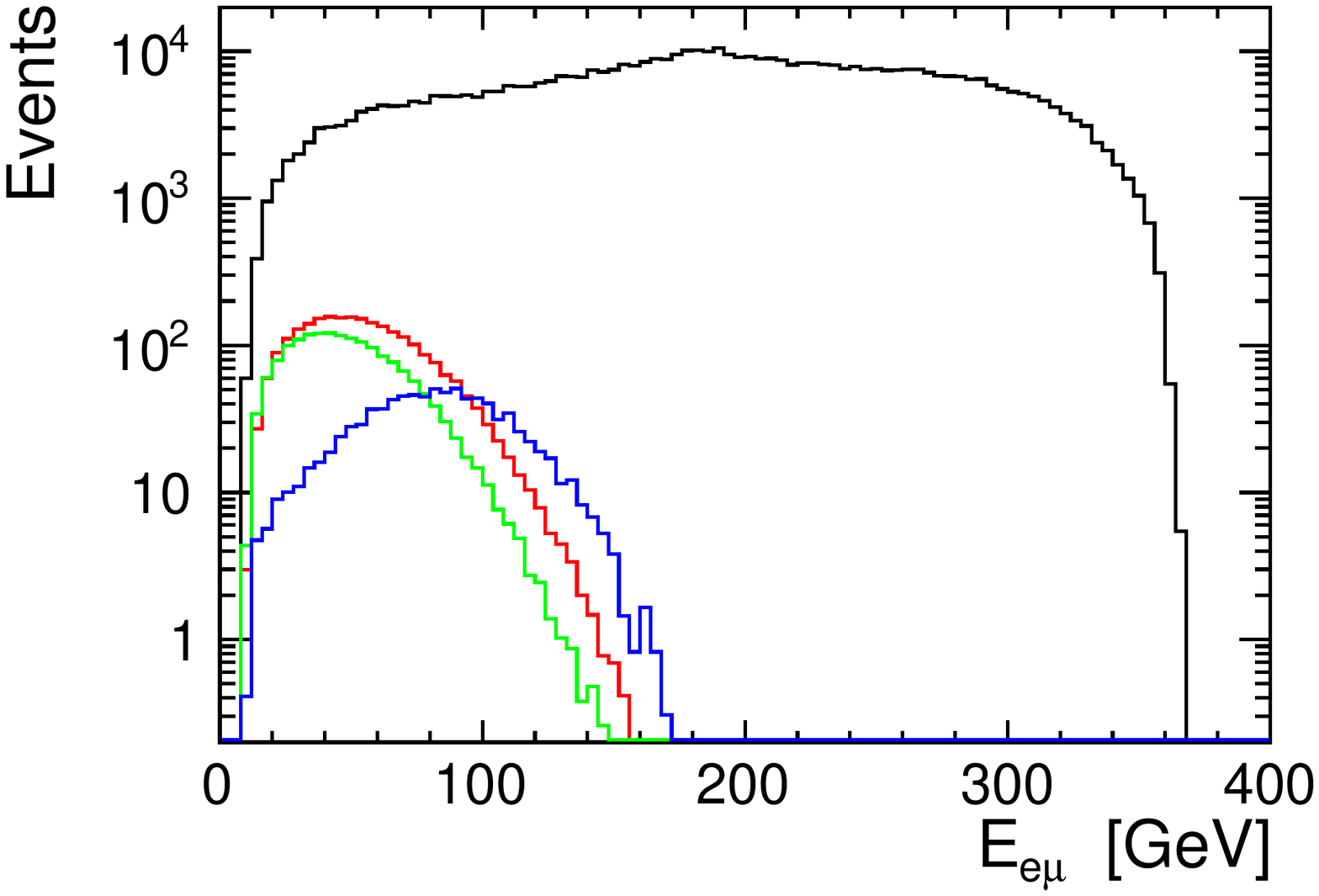}
  \includegraphics[width=0.49\textwidth]{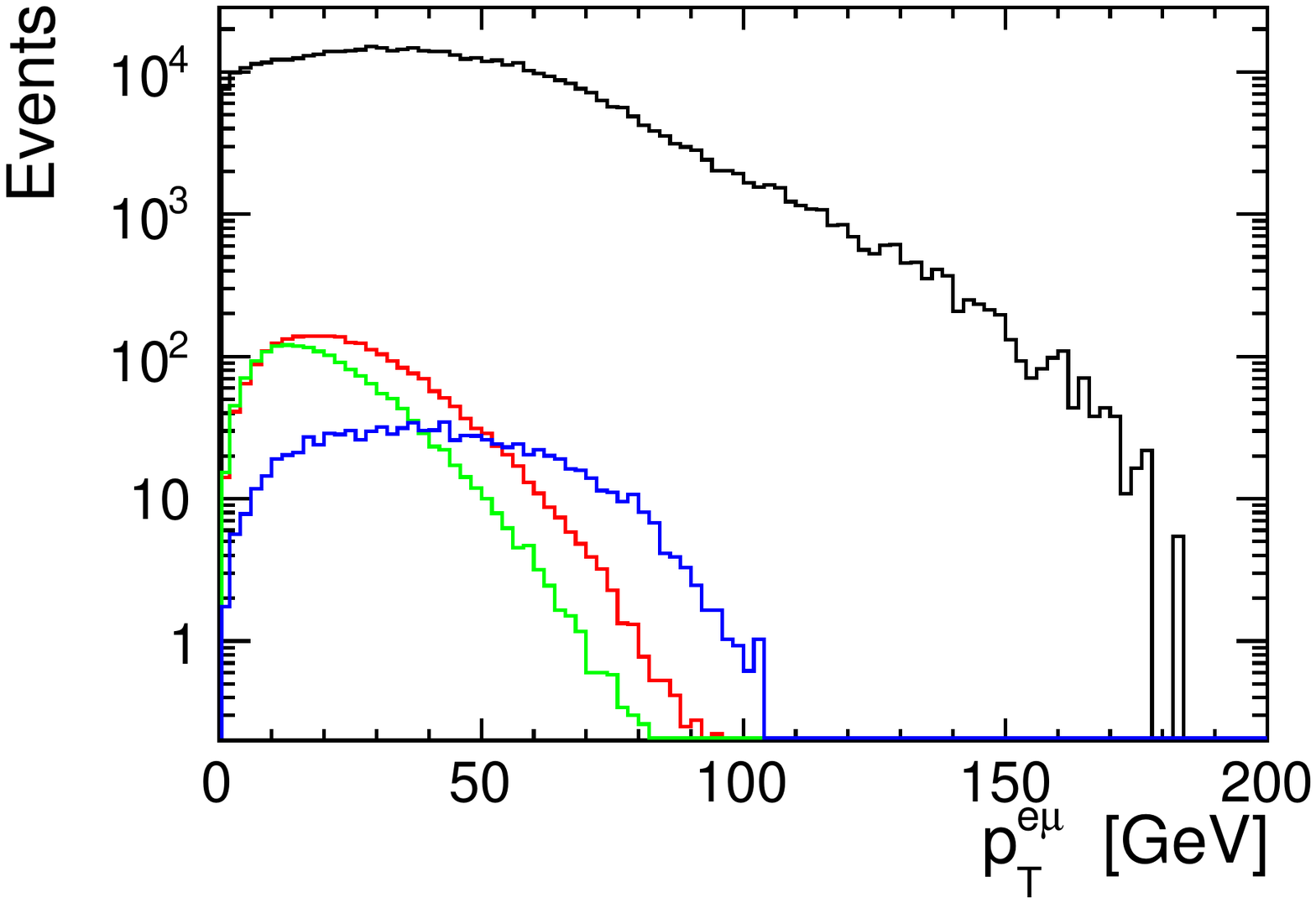}
  \includegraphics[width=0.49\textwidth]{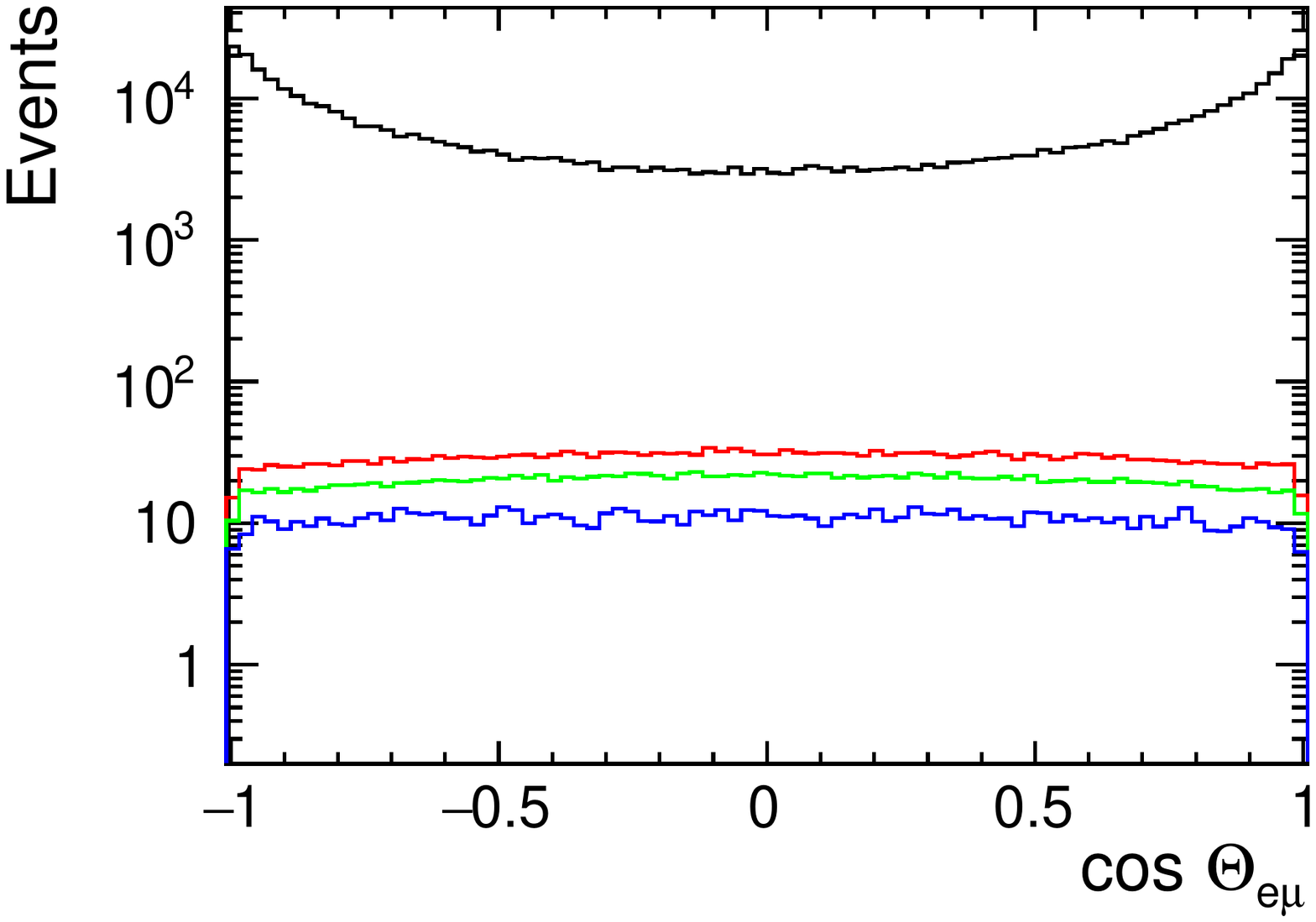}
  \includegraphics[width=0.49\textwidth]{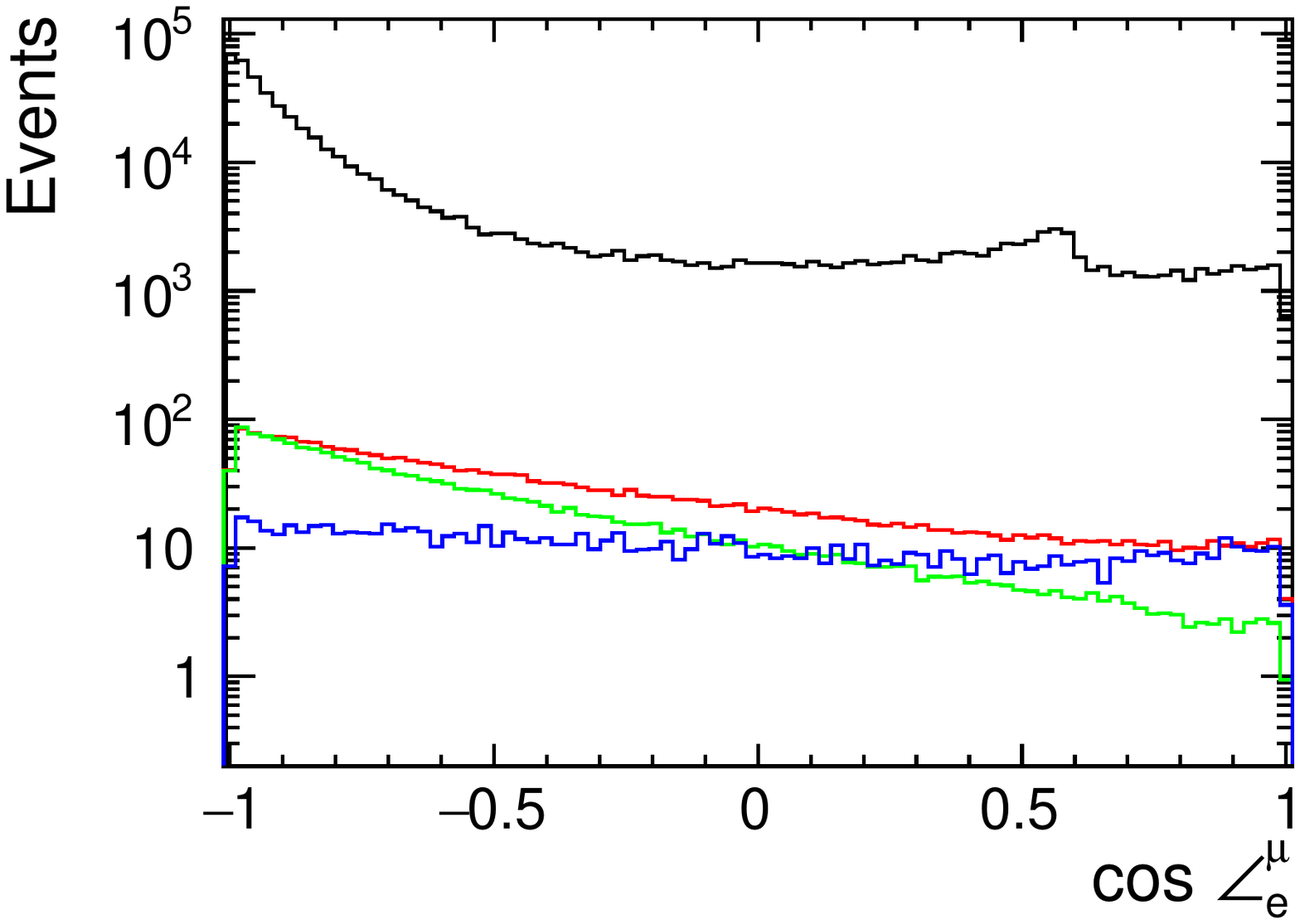}
\end{center}  
\caption{Distributions of the kinematic variables describing the
  leptonic final state considered in $H^+H^-$ analysis: lepton pair
invariant mass, M$_{e\mu}$, total longitudinal momentum, P$^{e\mu}_\text{Z}$, 
lepton pair energy, E$_{e\mu}$, total transverse momentum, P$^{e\mu}_\text{T}$,
pair production angle, $\Theta_{e\mu}$ and the angular distance
between the two leptons, $\cos\angle_{e}^{\mu}$.
Expected distributions for BP1 (red histogram), BP3 (green) and BP6
(blue) are compared with expected background (black
histogram). Samples simulated for CLIC running at 380\GeV are
normalised to 1\abinv. } 
\label{fig_plotsHpHm}
\end{figure}
However, kinematic distributions are very different, as two massive
scalars are produced in the signal case, reducing the kinematic space
available for lepton pair production.
In  \cref{fig_plotsHpHm} distributions of the selected variables
describing the leptonic final state for three benchmark scenarios
(BP1, BP3 and BP6) are compared with Standard Model expectations.
Clear differences between the signal and background distributions allow for 
efficient selection of signal-enhanced sample of events using the
multivariate analysis.
We follow the same procedure and the same set of input variables is used
as for the $AH$ analysis described above.
The BDT classification algorithm is trained separately for each
benchmark point to discriminate between signal and background
processes.
Examples of the BDT response distributions for the BP1 signal sample and
SM background samples simulated for 1\abinv at 380\GeV CLIC are shown
in \cref{fig_bdtHpHm}.
While due to a large SM background it is not possible to select the signal-dominated sample based
on the BDT response, 
the highest significance is obtained when selecting events with BDT response
above 0.12. About 1700 signal events are expected in the final sample (BDT
selection efficiency of 70\%) with background contribution of about
8500 events (BDT selection efficiency of 1.7\%), resulting in the
significance of the observation of about 17$\sigma$.

\begin{figure}[tb]
\begin{center}
  \includegraphics[height=0.35\textwidth]{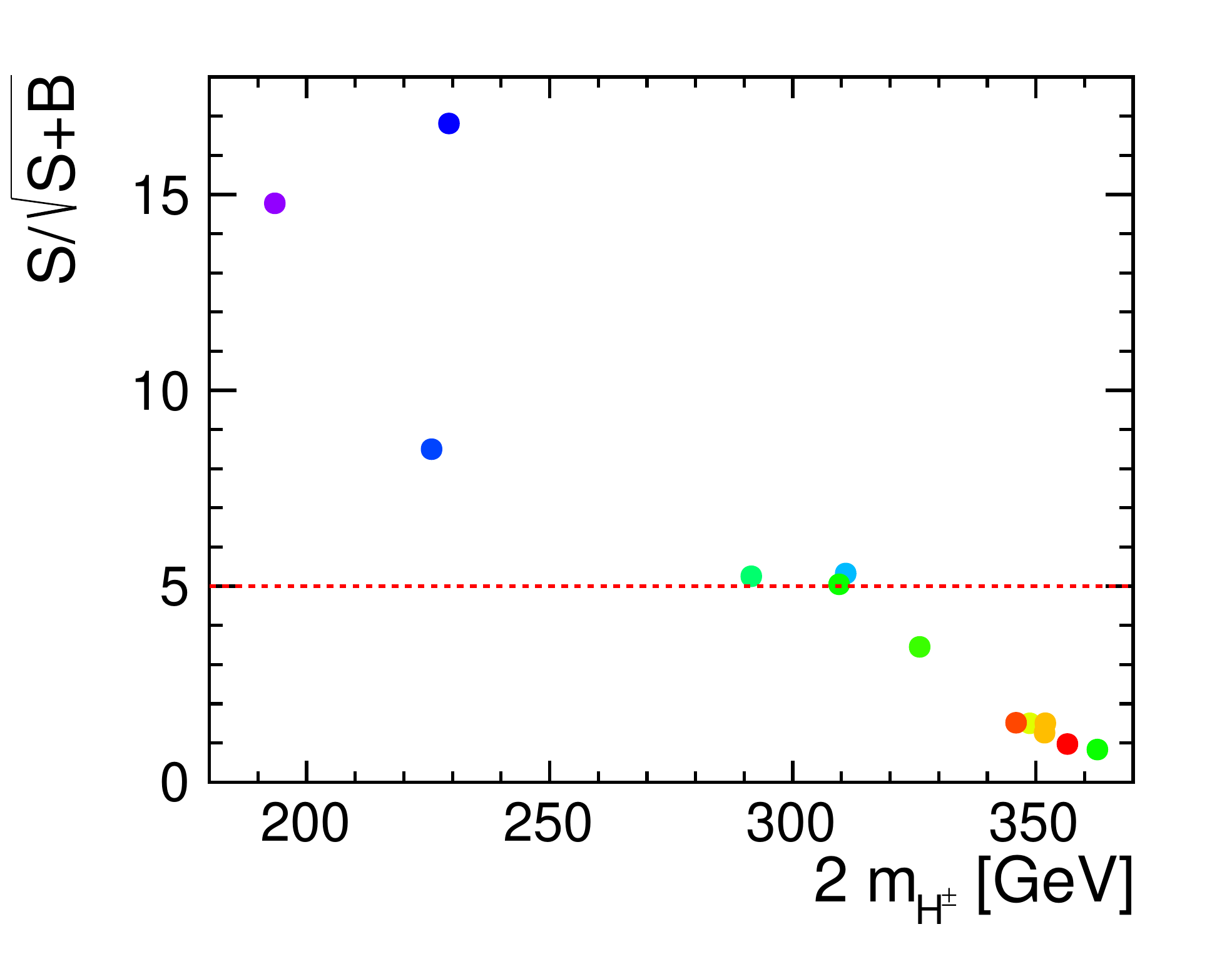}
  \includegraphics[height=0.35\textwidth]{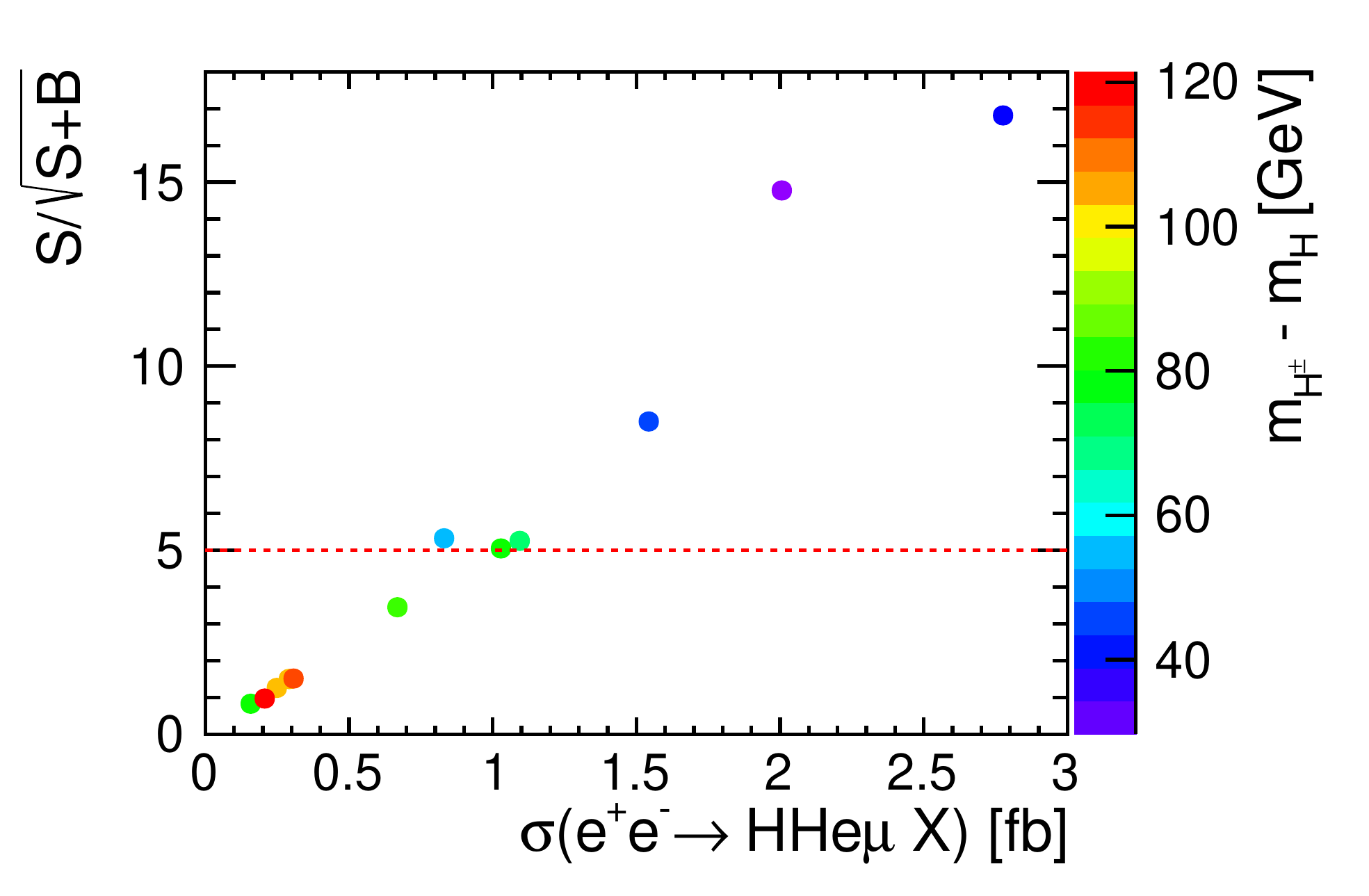}
\end{center}  
\caption{Expected significance of the deviations from the
  Standard Model predictions observed at 380\GeV CLIC for events with
  electron-muon pair in the final state ($e^+\mu^-$ or $\mu^+e^-$) as a function of twice the
  charged scalar mass {\sl (left)} and the production cross-section 
  for the considered signal channel {\sl (right)}, for different IDM
  benchmark points.
   Color indicates the mass splitting between the H$^\pm$ and H scalars 
  (right scale applies to both plots).
} 
\label{fig_sigHpHm}
\end{figure}
As was the case for the $AH$ channel, the expected significance of the
$e\mu$ signal is mainly related to the production cross-section for
the considered channel.
This is shown in \cref{fig_sigHpHm}, where the expected significance
for the electron-muon final state ($e^+\mu^-$ or $\mu^+e^-$) are
plotted as a function of $2m_{H^\pm}$ (left panel) and the production cross-section 
(right panel), for different IDM benchmark points.
Discovery at the initial stage of CLIC is only possible for scenarios
with signal cross-sections (in the electron-muon channel)
above about 1\fb.
This corresponds to charged scalar masses below roughly 150\GeV.
We do not observe any sizable dependence of the expected significance on the mass splitting between the charged and neutral inert scalars, $m_{H^{\pm}} - m_{H}$ (indicated by colour scale in \cref{fig_sigHpHm}), within the considered range of parameters.
Reduced signal channel cross section and thus reduced signal sensitivity 
observed for one of the benchmark points in  \cref{fig_sigHpHm} (BP2 with $m_{H^{\pm}}=112.8$\,GeV) is due to the significant contribution of cascade
decays, $H^\pm \to W^{\pm\star} A \to W^{\pm\star} Z^\star H$, 
which were not considered in the signal event selection.



\section{Inert Scalars at high-energy stages of CLIC}
\label{High energy}

We now turn to the discovery prospects of the two high-energy stages
at 1.5\TeV and 3\TeV with assumed integrated
luminosities of 2.5\abinv and 5\abinv \cite{Robson:2018zje}.
The same analysis procedure described in \cref{Low energy} was applied 
to signal and background samples simulated for high-energy CLIC
stages.
As before, proper energy spectra for CLIC \cite{Linssen:2012hp},
 based on
detailed beam simulations, were taken into account, which is 
crucial for a correct description of signal and background at high
collider energies. We applied the same generator-level cuts as before,
but did not make use of any additional pre-selection cuts. 
Furthermore, we extend our study to include additional high-mass
benchmark points not accessible 
at 380\GeV; these are listed in \cref{tab:bench2}.

\begin{figure}[tb]
\begin{center}
  \includegraphics[width=0.49\textwidth]{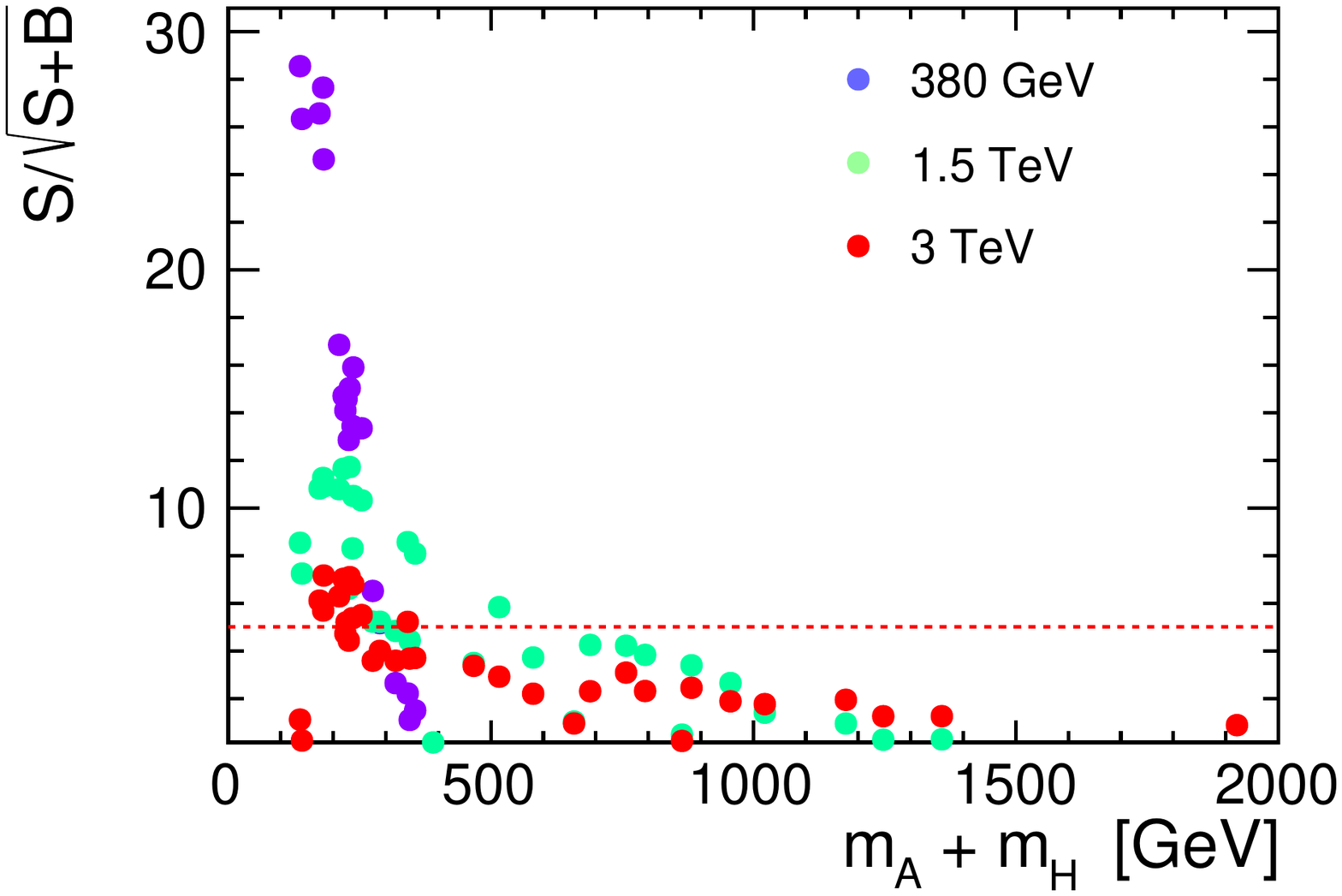}
  \includegraphics[width=0.49\textwidth]{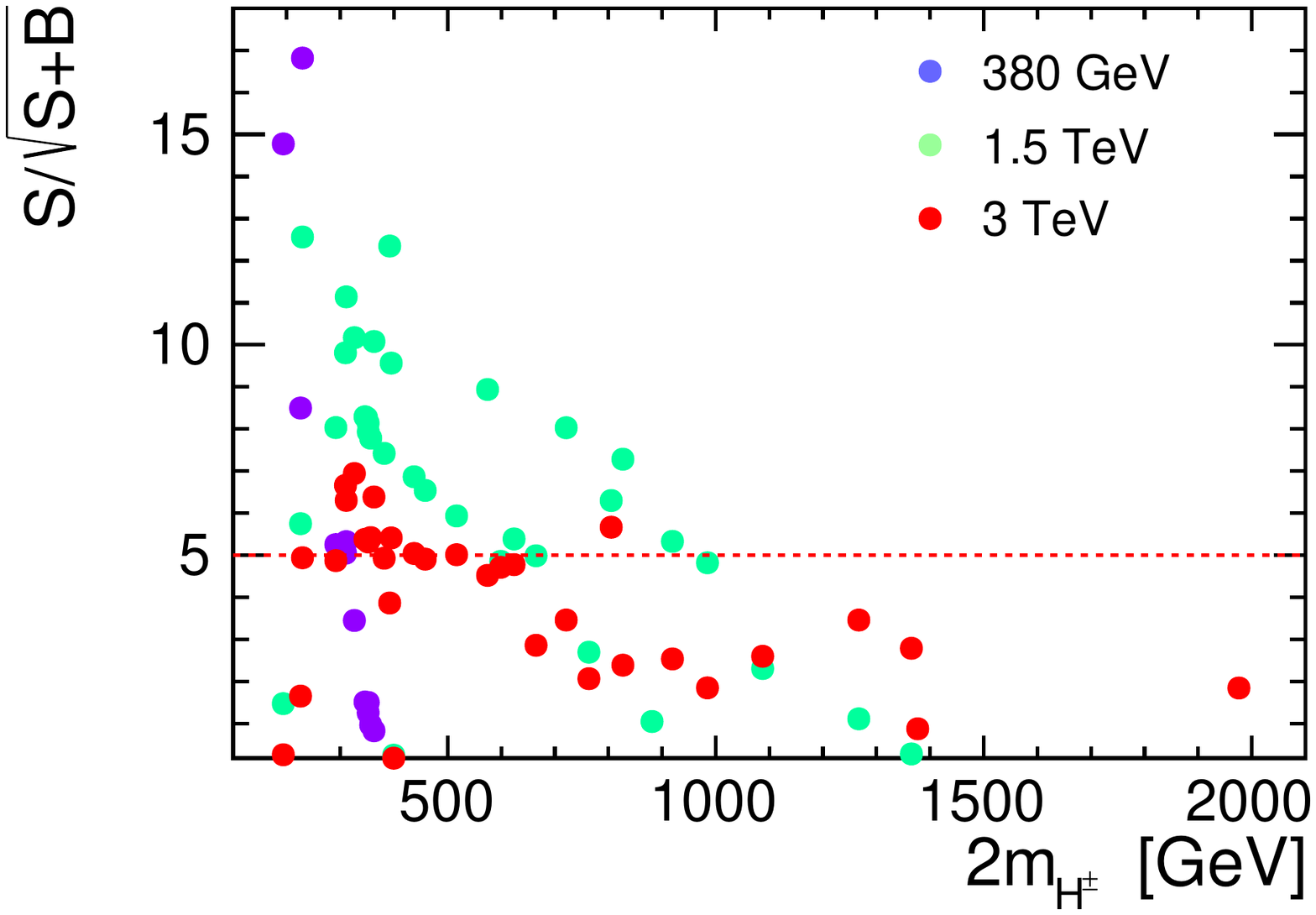}
\end{center}  
\caption{Significance of the deviations from the Standard Model
  predictions expected at the subsequent CLIC stages for: {\sl (left)}
  events with two muons in the final state ($\mu^+\mu^-$) as a
  function of the sum of neutral inert scalar masses and {\sl (right)}
  events with an electron and a muon in the final state ($e^+\mu^-$ or
  $e^-\mu^+$) as a function of twice the charged scalar mass, for
  IDM benchmark points in tables \ref{tab:bench} and \ref{tab:bench2}.}  
\label{fig_sigMass}
\end{figure}
In \cref{fig_sigMass}, we display the expected significances of the IDM
signal in the $AH$ and $H^+H^-$ channel as a function of the
inert scalar masses for subsequent CLIC running stages.
For \htb{the} $AH$ channel (muon-pair production), increasing the running energy
and integrated luminosity results in only a moderate extension of the
discovery potential of CLIC. With 2.5\abinv at 1.5\TeV scenarios
with the sum of neutral inert scalar masses up to about 550\GeV can be probed,
compared to about 290\GeV for 380\GeV running.
Prospects for high-energy CLIC running look significantly better if
the $H^+H^-$ production with the electron-muon final state is
considered.
Here the expected signal significance decreases much slower with the
charged scalar mass and we can probe masses up to about 500\GeV at
1.5\TeV, compared to 150\GeV at the first CLIC stage (see
\cref{fig_sigMass} right panel).

\begin{figure}[tb]
\begin{center}
  \includegraphics[width=0.49\textwidth]{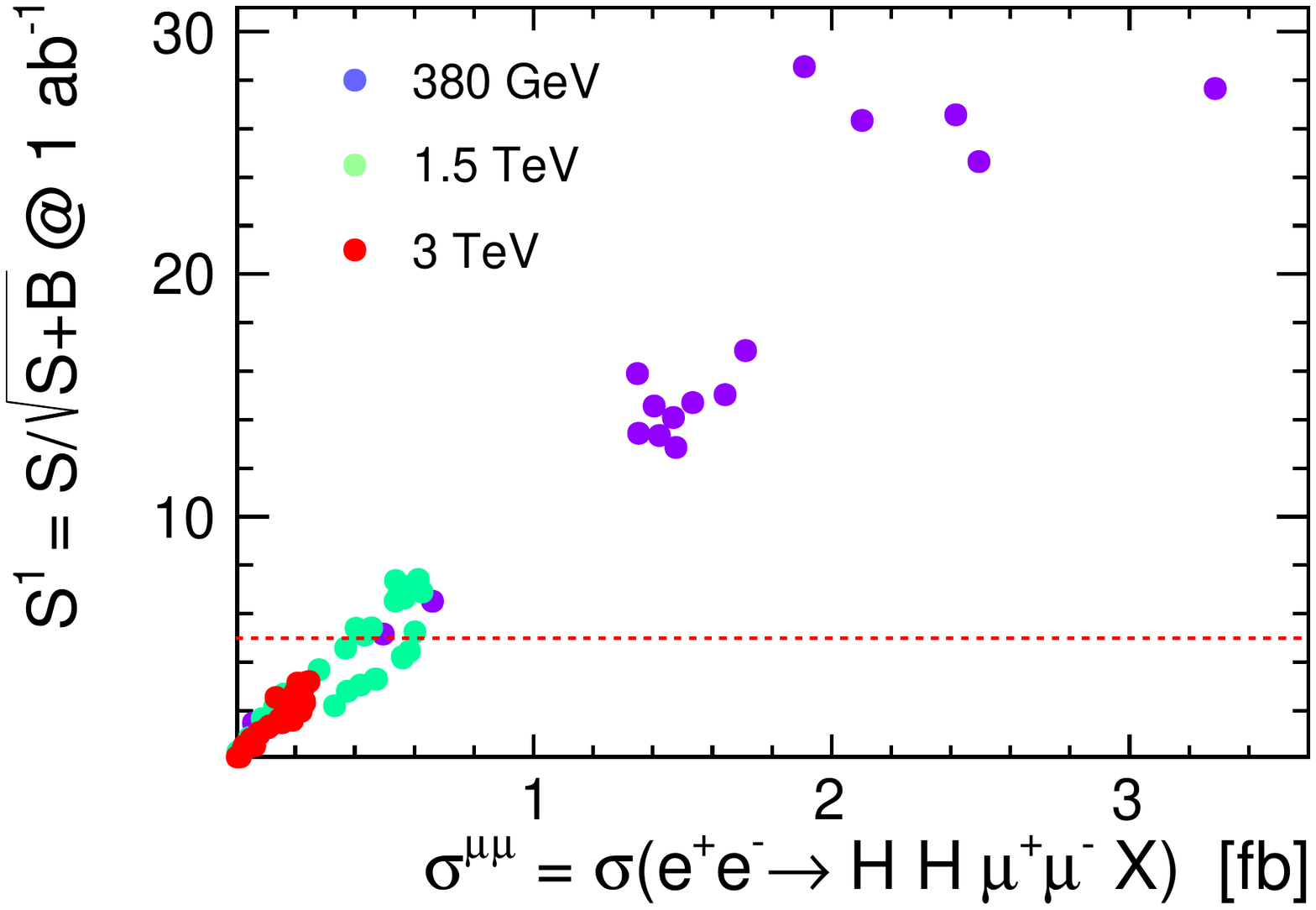}
  \includegraphics[width=0.49\textwidth]{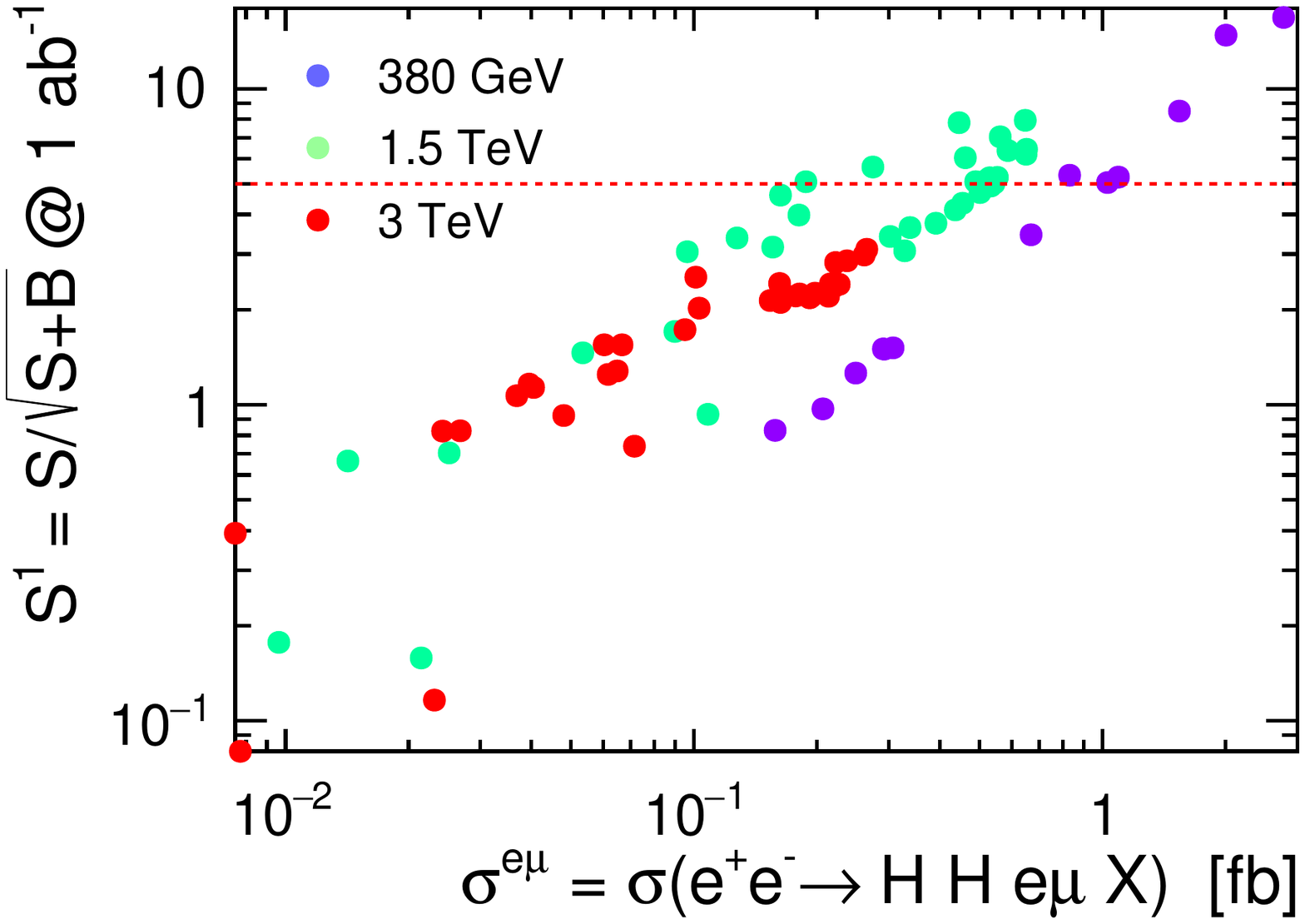}
\end{center}  
\caption{Significance of the deviations from the Standard Model
  predictions expected at different CLIC running stages, assuming the
  same integrated luminosity of 1\abinv, as a function of the
  signal cross-section in the considered channel, for: {\sl (left)} events with 
  two muons in the final state ($\mu^+\mu^-$) and {\sl (right)} events with
  electron-muon pair production ($e^+\mu^-$ or $e^-\mu^+$), for the
  IDM benchmark points in tables \ref{tab:bench} and
  \ref{tab:bench2}.}
\label{fig_sigCross}
\end{figure}

 The significance is mainly driven by the signal production
cross section and is approximately proportional to the
square-root of the integrated luminosity. For parameter points that
are already accessible at Stage 1 the $AH$ production cross sections
decrease with the collision energy much faster than most of the
backgrounds and the significance of observation decreases at Stage~2.
Only for points with $m_A+m_H\,\gtrsim\,300\,\GeV$, which are close
to the production threshold at Stage 1, higher integrated luminosity and
the production cross sections enhanced by up to a factor of 2 result in better
sensitivity at center-of-mass energy of 1.5\,TeV.
Similarly, when going from 1.5\,TeV to 3\,TeV, the significance of
observation increases only for scenarios with $m_A+m_H\,\gtrsim\,1.2\TeV$.

As we search for the signal contribution on top of a much larger
background, we expect that the significance is (to a first
approximation) proportional to the square-root of the integrated
luminosity.
In order to compare the CLIC sensitivity to the IDM benchmark scenarios at different
energies, we scale the expected significance at high-energy stages to
the integrated luminosity of 1\abinv assumed for 380\GeV
running. This allows us to separate luminosity and cross-section
contributions to the overall significance, and will also allow for
projections of the discovery potential at arbitrary luminosities.

In \cref{fig_sigCross}, we show the scaled significance results, presented as a function of the signal
production cross-section. 
For the $AH$ channel, which leads to $\mu^+\mu^-$ final states, a universal linear
dependence on the signal cross-section is observed which does not seem
to depend on the running energy.
Significant (above 5$\sigma$) observation is possible for cross-sections 
roughly larger than  0.5\,\fb\ (for higher luminosities, these
should be rescaled by $\sqrt{1\,/{\cal L}\cdot \ab}$).

For the $H^+H^-$ channel, however, leading to $e^\pm \mu^\mp$ final states, the high-energy
running of CLIC clearly gives better sensitivity to  heavy IDM scenarios
(for points with similar production
cross-section and assuming same luminosity) than the
initial CLIC stage (see \cref{fig_sigCross} right). The relatively
large differences between different BPs with similar cross-sections at
the same center-of-mass energies
originate from the mass scale of the heavy scalars.

\begin{figure}[tb]
\begin{center}
  \includegraphics[width=0.49\textwidth]{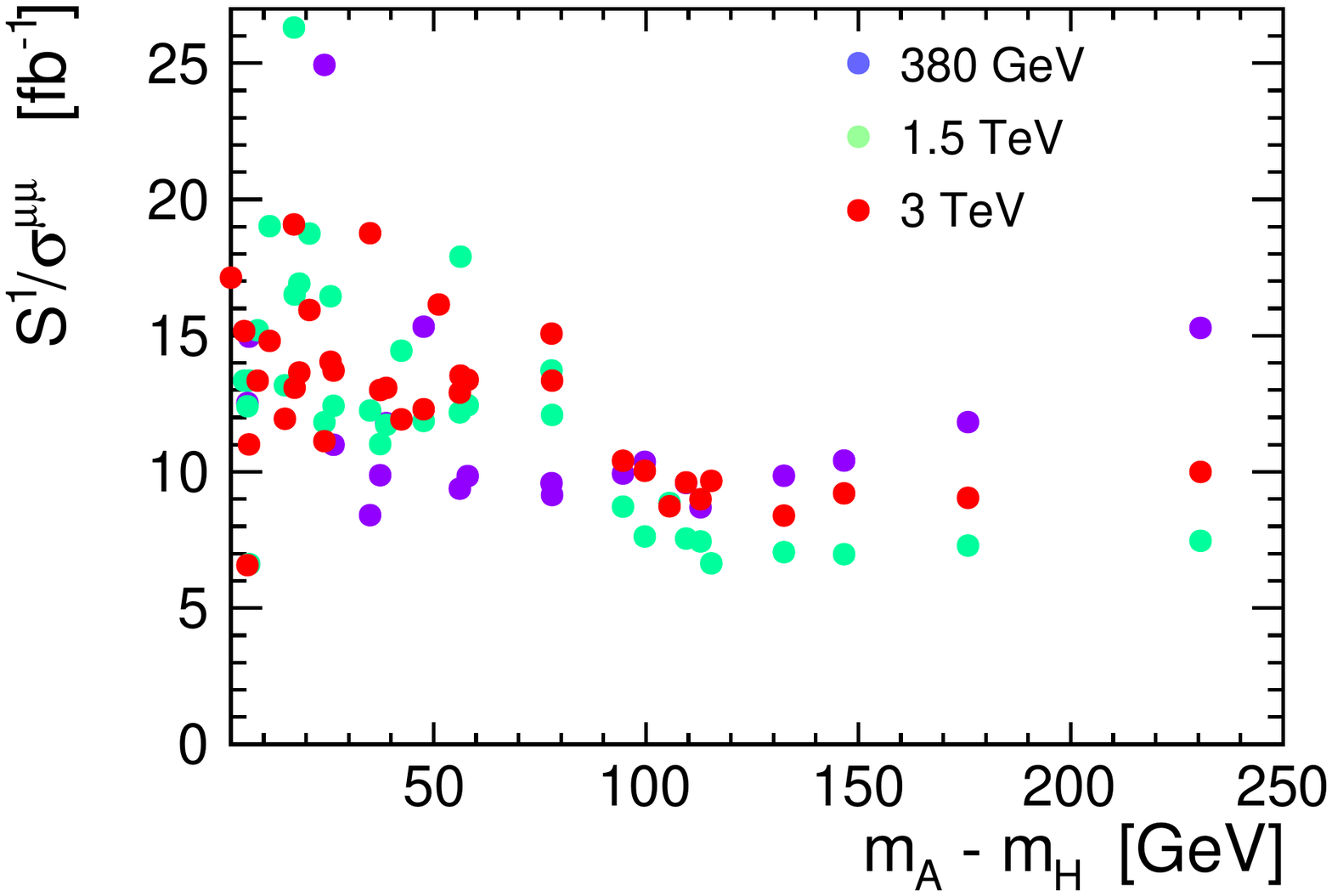}
  \includegraphics[width=0.49\textwidth]{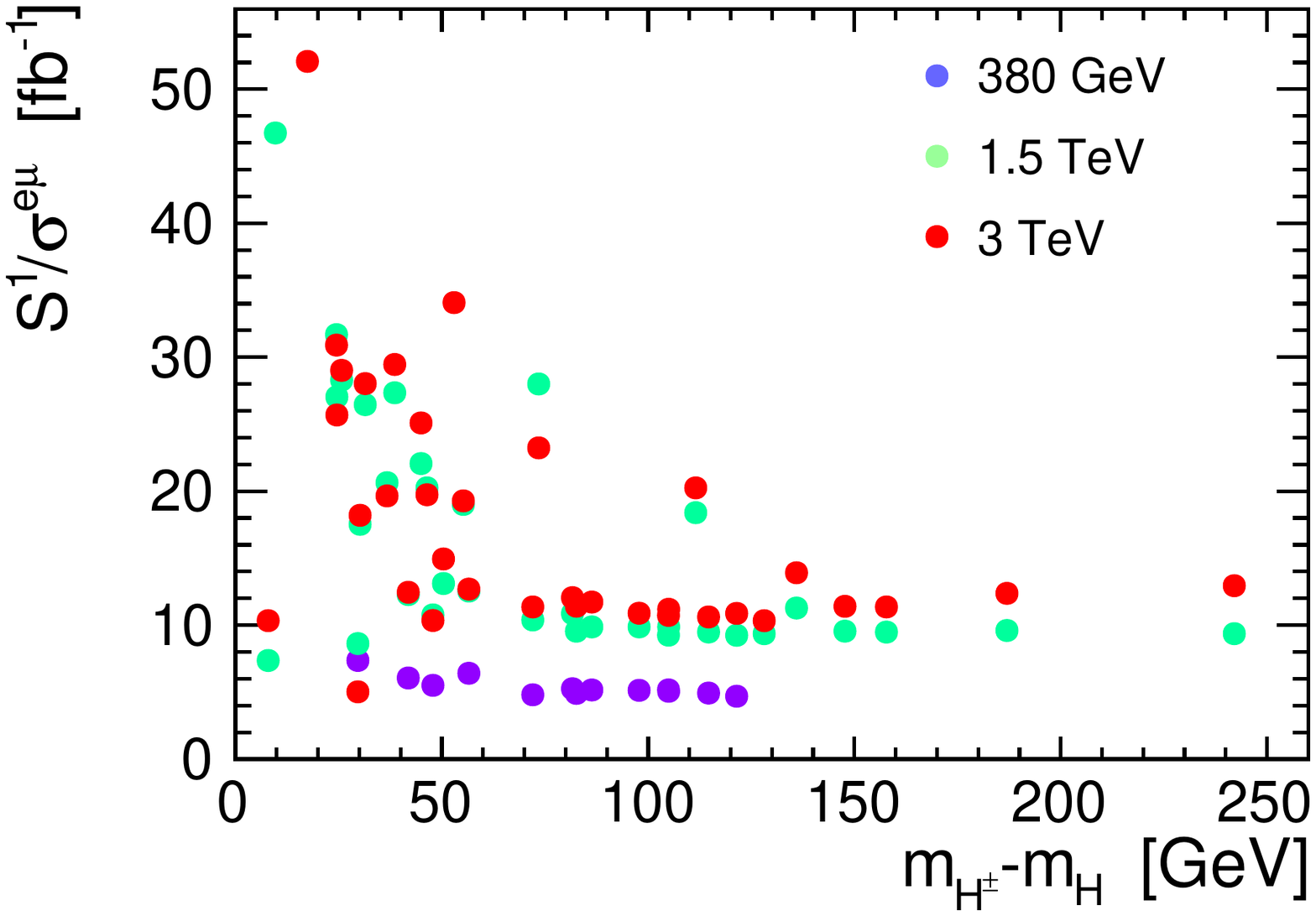}
\end{center}  
\caption{Ratio of the expected significance scaled to the integrated
  luminosity of 1\abinv to the signal cross-section in the
  considered channel: {\sl (left)} with two muons in the final state
  ($\mu^+\mu^-$) and {\sl (right)} with electron-muon pair production
  ($e^+\mu^-$ or $e^-\mu^+$), as a function of the scalar mass
  differences, for different IDM  benchmark points.
}  
\label{fig_sig2cross}
\end{figure}
Finally, we investigate the dependence of the signal significance on
the mass difference between neutral/charged inert scalar and the DM
candidate. In \cref{fig_sig2cross}, we show the ratio
of the expected significance (scaled to the integrated  luminosity of
1\abinv) to the signal cross-section in the considered channel,
as a function of the corresponding scalar mass difference.
This ratio indicates the expected significance for the particular mass splitting, 
assuming the reference signal channel cross section of 1\,fb.
For  $AH$ production (muon-pair channel) at high-energy stages, the
experimental sensitivity seems to be significantly better for low mass
differences, below $m_{Z}$, when the virtual $Z$ boson is produced in
the $A$ boson decay, $A \rightarrow Z^{(\star)} H$.
This is because signal events can be better separated from the SM backgrounds 
for such scenarios.
One can also note that for high mass differences, $m_{A} - m_{H} > m_{Z}$, 
the experimental sensitivity is clearly better for low-energy running.

The situation is similar for the $H^+H^-$ production signal in the 
electron-muon channel.
For high running energies, a better sensitivity is expected for low mass
differences  when the virtual $W^\pm$ boson is produced in the charged
scalar decay.
However, it is also clear that the experimental sensitivity is much
better at high-energy running than at the first CLIC stage and this
observation does not depend on the considered dark scalar mass difference.

The results presented in \cref{fig_sigMass} seem to indicate that many
high mass IDM scenarios will remain inaccessible at CLIC, even at
high energies.
However, one has to stress that this is mainly due to the small branching ratios for the considered leptonic final states: 3.3\% for $AH\to HH \mu^+ \mu^-$ and 2.3\% for
$H^+H^- \to HH\mu^\pm e^\mp \nu\nu$.
For scenarios where the signal cross sections in the dilepton channel are too small, 
it might be worthwhile to investigate semi-leptonic decays 
in the $H^+H^-$ production channel.
Due to the much larger branching ratios (28.6\% for $H^+H^- \to HH \ell^\pm \nu qq'$,
with $\ell = e, \mu$)
the expected number of $H^+ H^-$ signal events in the semi-leptonic final state 
is over an order of magnitude larger than for the electron-muon signature.
As a similar scaling is expected for the background processes
(dominated by the $W^+W^-$ production), we expect that the
significance of the observation in the semi-leptonic channel should be
increased by at least a factor of 3 (corresponding to the ten-fold
increase of the integrated luminosity).
An investigation in this channel could furthermore profit from a full
reconstruction of the $W^\pm$ that decays hadronically, which allows to
use the reconstructed $W$ boson mass and energy as additional
discriminators in the BDT algorithm.
However, a proper estimate of the expected significance for this case
would require a much more detailed analysis, including parton showering,
hadronisation and detector response simulation, and event reconstruction 
with particle flow algorithm and final state
reconstruction using accurate jet clustering and lepton identification
processes. This is beyond the scope of the work presented here.
%



\section{Conclusions}
\label{Conclusion}

In this work, we have studied prospects for discovery of inert scalars
of the  Inert Doublet Model at CLIC running at 380\,\GeV,
1.5\,\TeV~ and 3\,\TeV.
A set of benchmark points, proposed in \cite{Kalinowski:2018ylg} and satisfying all
experimental and theoretical constraints, has been considered.
We focused on pair-production of charged dark scalars $H^+\,H^-$ and
production of the DM candidate with the second neutral scalar
boson, $HA$, with subsequent decays to leptonic final states.
Signal and background event samples were generated with {\tt
WHizard~2.2.8}, taking into account all processes that lead to the 
considered final states. 
Signatures for production of new scalars were searched for in the
kinematic distributions for events with exclusive production of two
muons or an electron and a muon.
Significance of the possible observation was studied using
multivariate analysis methods.

We found that most of the low-mass benchmark scenarios proposed in
\cite{Kalinowski:2018ylg} can be observed with high significance in the
di-muon channel already with 1\abinv collected at 380\GeV
(the first stage of CLIC), provided that the sum of neutral inert
scalar masses, $m_A + m_H < 290$\GeV.  
Similar constraints also apply to the observation of the charged
scalar pair-production (electron-muon pair-production channel), which
is however fulfilled for fewer scenarios. 

Scenarios which are not kinematically accessible at the first stage of
CLIC can be searched for at high-energy stages, at 1.5\,TeV and 3\,TeV.
The signal production cross-section for both considered
channels decreases significantly with energy, much faster than for the
corresponding background processes.
Signal cross sections for the considered final states are further reduced 
by the small branching fractions for the dilepton channels.
We found that at 1.5\,TeV the discovery reach is extended to the sum of scalar masses 
of about 550\GeV in the dimuon channel
and for charged scalar masses up to about 500\GeV in the
$e^\pm\,\mu^\mp$ channels. 
For the scenarios considered here, increasing the center-of-mass
energy to 3\TeV does not significantly improve the sensitivity.
Therefore, the observation of the inert scalar production in the
leptonic channels will be challenging at high-energy CLIC and
a significant signal is only expected for relatively low masses.
However, higher significance and the discovery reach extending 
up to the kinematic limit could be expected for $H^+ H^-$
production in the semi-leptonic final state (isolated lepton and a
pair of jets or one massive jet). This is in the line of future work.
%


\section*{Acknowledgements}

This research was supported in parts by the National Science Centre,
Poland, the HARMONIA project under contract UMO-2015/18/M/ST2/00518
(2016-2019) and OPUS project under contract UMO-2017/25/B/ST2/00496
(2018-2021), as well as the COST Action CA15108. TR was supported by Michigan State University through computational resources
provided by the Institute for Cyber-Enabled Research, by grant K 125105 of
the National Research, Development and Innovation Fund in Hungary, and
by the European Union through the European Regional Development Fund - the Competitiveness and Cohesion Operational Programme (KK.01.1.1.06). 
TR also thanks the Galileo Galilei Institute for Theoretical Physics for the hospitality 
and the INFN for partial support during the completion of this work. The work of WK was partially supported by the German Research Foundation (DFG) under grants
number STO 876/4-1 and STO 876/2-2.
JK and WK  thank Gudrid Moortgat-Pick for her hospitality and the DFG for a partial support 
through the SFB~676 ``Particles, Strings and the Early Universe'' 
during the initial stage of this project.


\bibliographystyle{JHEP}

\bibliography{idm_yrep_litp}

\end{document}